\documentclass[reprint, superscriptaddress,
 amsmath,amssymb,
 aps,pra]{revtex4-2}

\usepackage{graphicx}
\usepackage[english]{babel}
\usepackage[utf8x]{inputenc}
\usepackage{dcolumn}
\usepackage[colorinlistoftodos]{todonotes}
\usepackage{xcolor}
\usepackage{mathtools}
\usepackage{bm}
\usepackage{hyperref}
\usepackage{float}

\usepackage{psfrag}
\usepackage{indentfirst} 
\usepackage{url}
\usepackage{enumitem}
\usepackage{tensor}
\usepackage{braket}
\usepackage{bbm}
\usepackage{dsfont}
\usepackage{amsfonts,latexsym,cancel}
\usepackage{rawfonts}
\usepackage{pictexwd}
\usepackage{bbold}
\usepackage{wrapfig}
\usepackage{subfigure} 
\usepackage{soul}

\DeclareMathOperator{\Tr}{Tr}

\usetikzlibrary{arrows}

\begin{document}

\preprint{}

\title{Classical analogs of generalized purities, entropies, and logarithmic negativity}

\author{Bogar Díaz}
\email{bodiazj@math.uc3m.es}
\affiliation{Departamento de Matem\'aticas, Universidad Carlos III de Madrid, Avenida  de la Universidad 30, 28911 Legan\'es, Spain} 
\affiliation{Grupo de Teor\'ias de Campos y F\'isica Estad\'istica. Instituto Gregorio Mill\'an (UC3M), Unidad Asociada al Instituto de Estructura de la Materia, CSIC, Serrano 123, 28006 Madrid, Spain}
\affiliation{Instituto de Ciencias Matem\'aticas (ICMAT), CSIC-UAM-UC3M-UCM, C. Nicol\'as Cabrera 13-15, 28049 Madrid, Spain}
\author{Diego Gonzalez}
\email{dgonzalezv@ipn.mx}
\affiliation{Escuela Superior de Ingenier\'ia Mec\'anica y El\'ectrica - Instituto Polit\'ecnico Nacional (ESIME - IPN), 07738, CDMX, M\'exico} 
\author{Marcos J. Hern\'andez}
\email{mhm@ciencias.unam.mx}
\affiliation{Departamento de F\'isica de Altas Energ\'ias, Instituto de Ciencias Nucleares, Universidad Nacional Aut\'onoma de M\'exico, Apartado Postal 70-543, Ciudad de M\'exico, 04510, M\'exico} 
\author{J. David Vergara}
\email{vergara@nucleares.unam.mx}
\affiliation{Departamento de F\'isica de Altas Energ\'ias, Instituto de Ciencias Nucleares, Universidad Nacional Aut\'onoma de M\'exico, Apartado Postal 70-543, Ciudad de M\'exico, 04510, M\'exico}

\begin{abstract}
It has recently been proposed classical analogs of the purity, linear quantum entropy, and von Neumann entropy for classical integrable systems, when the corresponding quantum system is in a Gaussian state. We generalized these results by providing classical analogs of the generalized purities, Bastiaans-Tsallis entropies, R\'enyi entropies, and logarithmic negativity for classical integrable systems. These classical analogs are entirely  characterized by the classical covariance matrix. We compute these classical analogs exactly in the cases of linearly coupled harmonic oscillators, a generalized harmonic oscillator chain, and a one-dimensional circular lattice of oscillators. In all of these systems, the classical analogs reproduce the results of their quantum counterparts whenever the system is in a Gaussian state. In this context, our results show that quantum information of Gaussian states can be reproduced by classical information.
\end{abstract}

\maketitle

\section{Introduction}\label{sec_introduction}


In recent times, many measures of quantum effects have emerged; among them, we can highlight the covariance matrix \cite{Amari}, purity \cite{Jaeger2007}, von Neumann entropy \cite{vonN}, R\'enyi entropies \cite{renyi1970}, Bastiaans-Tsallis entropies \cite{Bastiaans84, Tsallis88}, logarithmic negativity \cite{Vidal2002}, Berry phase \cite{Berry1984}, and quantum geometric tensor \cite{Provost}. Each of these quantities  has its utility since they easily discern specific characteristics of quantum phenomena. For instance, the von Neumann entropy encodes the degree of mixing of a quantum state. Also, the R\'enyi entropies are the natural generalization of the von Neumann entropy by deforming it into a parametrized entropy by a positive real number $\alpha$, which characterizes different aspects of the entanglement spectrum in a similar way to higher moments of a probability distribution. Furthermore, the R\'enyi entropies have also found various applications \cite{Muller}. In particular, these entropies are widely used as a technical tool in information theory\cite{Amari}, statistical mechanics\cite{Lenzi}, and recently using the maximum entropy principle applied to the R\'enyi entropies, a generalized formulation of quantum thermodynamics was developed \cite{Kim2018}. In addition, the Renyi entropies have been essential to defining multifractality phenomena \cite{Jizba}. On the other hand, the Bastiaans-Tsallis entropies were introduced by Bastiaans in the context of quantum optics \cite{Bastiaans84} and, independently, by Tsallis in statistical mechanics \cite{Tsallis88}. These entropies quantify the degree of mixedness of a state by the amount of information it lacks, and it is a nonadditive quantity. Also, it  is a fundamental tool to describe nonextensive Statistical Mechanics \cite{Tsallisb}, which has managed to describe a good number of phenomena ranging from optical phenomena such as anomalous transport in an optical lattice \cite{Tsallis1, Tsallis2} to heavy-ion collisions \cite{Tsallis4, Tsallis5}. Other applications of the Tsallis entropies involve recognizing that the parameter $\alpha=1/2$ is a natural operational measure of nonclassicality \cite{Carmi}. Another interesting measure of entanglement is logarithmic negativity \cite{Vidal2002}, one of the most practical measures of quantifying mixed entanglement \cite{Horode}, which can be used as an upper bound on distillable entanglement\cite{Audenaert2} and has an unambiguous operational interpretation \cite{Audenaert2002}. Furthermore, it allows a more varied selection of subsystems to be analyzed concerning entanglement with each other and through a medium. However, the logarithmic negativity is not convex, implying it can increase under mixing. 

On the other hand, for some years, there have been attempts to construct classical analogs of quantum quantities, such as the Berry phase \cite{Hannay_1985, Chaturvedi1987,Zhang}, linear quantum entropy \cite{Gong2003}, \ 
quantum geometric tensor \cite{Gonzales2019,Alvarez2019,Gonzalez2020} and more recently the quantum covariance matrix, purity, and  von Neumann entropy \cite{DGGV2022}. These constructions have had the purpose of discovering to what extent quantum effects can be computed using classical mechanics tools and if these tools might shed some light on new properties of these quantum effects. In addition, these classical analogs aim to see if they can be more easily calculated than their quantum counterparts, at least in some particular cases. What has been discovered is that, in the case of Gaussian states, the classical analogs describe most of the quantum effects with remarkable accuracy, although some minor differences can be understood in terms of ordering anomalies \cite{Alvarez2019}. In this article, we want to extend and compare these results in the framework of classical integrable systems by introducing classical analogs of the generalized purities, R\'enyi entropies, Bastiaans-Tsallis entropies, and logarithmic negativity. We shall see that this provides a new and alternative approach that effectively describes these other fundamental quantities of the quantum spectrum.
Furthermore, some quantum mechanical systems have non-perturbative effects, which can be described with kink-like solutions\cite{kinks}, instanton-like solutions \cite{instanton}, flow tubes\cite{flow}, domain walls\cite{Domainwall}, and others. The solution to these systems generally starts by obtaining an exact classical solution of a Euclidean version of the problem. So our idea of considering classical attack methods could give us information on a non-perturbative nature of a quantum system, which perhaps cannot be obtained by standard perturbative procedures.

The paper is organized as follows. In Sec. \ref{secprelimi}, we fix the notation used throughout the paper and give a brief  summary of the quantum purity, linear quantum entropy, and  von Neumann entropy, as well as of their classical analogs for classical integrable systems introduced in \cite{DGGV2022}. We also include here an example that illustrates the use and  main features of these classical analogs. In Sec.~\ref{sec3}, we review some basics about the generalized purities, Bastiaans-Tsallis, R\'enyi entropies, and logarithmic negativity. A key feature for our purposes is that these functions are entirely determined by the quantum covariance matrix if the system is in a Gaussian state. Then, taking this into account, in Sec. \ref{sec4}, we introduce classical analogs of generalized purities, Bastiaans-Tsallis entropies, R\'enyi entropies, and logarithmic negativity for classical integrable systems. Analogously to their quantum counterparts, these classical functions are completely determined by the classical covariance matrix of the classical integral system. In Sec.\ref{examples}, we provide three examples to illustrate the application of the introduced classical analogs, confronting the results with their quantum counterparts. The first example considers a system of linearly coupled harmonic oscillators, the second one a generalized harmonic oscillator chain, and the third one a one-dimensional circular lattice of oscillators. In all of these systems, the classical analogs agree with their quantum versions, for Gaussian states. Finally, in Sec. \ref{conclusions}, we give our conclusions and some comments.

\section{Preliminaries}\label{secprelimi}

We begin by considering a quantum system described by a Hamiltonian $\hat{\bf{H}}(\hat{\bf{q}},\hat{\bf{p}})$ with a set of phase space operators, $\hat{\bf{q}}=\{\hat{\bf{q}}^{a}\}$ and $\hat{\bf{p}}=\{\hat{\bf{p}}_{a}\}$ ($a=1,\dots, N$). We use bold letters with a hat to denote quantum operators. For a given quantum state $|m\rangle$,  the quantum covariance matrix $\boldsymbol{\sigma} = (\sigma_{\alpha \beta})$ is a $2N\times 2N$ matrix whose elements are
\begin{align}
\sigma_{\alpha \beta} = \frac{1}{2} \langle \hat{\bf{r}}_{\alpha} \hat{\bf{r}}_{\beta} + \hat{\bf{r}}_{\beta} \hat{\bf{r}}_{\alpha}  \rangle_{m}-\langle \hat{\bf{r}}_{\alpha}\rangle_{m} \langle \hat{\bf{r}}_{\beta} \rangle_{m}\,, \label{qumet}
\end{align}
where $\hat{\bf{r}}=\{\hat{\bf{r}}_{\alpha}\}=(\hat{\bf{q}}_1,\dots,\hat{\bf{q}}_N,\hat{\bf{p}}_1,\dots,\hat{\bf{p}}_N)^{\intercal}$ ($\alpha,\, \beta =1,\dots,2N$) is a $2N$-dimensional column vector  and $ \langle \hat{\bf{O}} \rangle_{m}=\langle m | \hat{\bf{O}} | m\rangle$ stands for the expectation value of an operator $\hat{\bf{O}}$. In the classical setting, the quantum covariance matrix has a counterpart for classical integrable systems \cite{DGGV2022}. Consider a classical integrable system defined by a Hamiltonian $H(q,p)$ with a set of  canonical coordinates and momenta, $q=\{q^{a}\}$ and $p=\{p_{a}\}$ ($a=1,\dots,N$). Within the Wigner function formalism, we consider the WKB approximation for the wave function $\psi_m(q)=\langle q |\psi_m\rangle$,  which reads
\begin{align}
    \psi_m(q)=\frac{1}{(2\pi)^N}\left|\det \frac{\partial^2S(q,I_m)}{\partial q_a \partial I_b}\right|e^{\tfrac{i}{\hbar}S(q,I_m)}\, ,
\end{align}
where $S(q,I_m)$ is the classical action corresponding to the particular torus $I_m$. Following Berry \cite{Berry1977}, we built the  semi-classical Wigner function 
\begin{align}
\label{wignersemi}
    W_m(q,p)=&\frac{1}{(2\pi^2\hbar^2)^{N}}\int dx \ e^{\tfrac{i}{\hbar}\left(S(q+x,I_m)-S(q-x,I_m) -2p\cdot x\right)}\nonumber \\
    &\left|\det \frac{\partial^2S(q+x,I_m)}{\partial q_a \partial I_b} \det \frac{\partial^2S(q-x,I_m)}{\partial q_a \partial I_b}\right|\, ,
\end{align}
by considering a classical limit taking only linear corrections in the exponential and performing a  canonical transformation of the phase-space coordinates to action-angle variables $I=\{I_a\}$ and $\varphi=\{\varphi_a\}$ ($a=1,\dots, N$). In this case, the Wigner function of the system is reduced to a delta function
\begin{align}
\label{wignerclas}
    W_m(q,p)=& \delta\left(I(q,p)-I_m\right),
\end{align}
involving only the action variables of the system  and the value of the action corresponding to the $m$ torus, i.e., $|\psi_m\rangle$, quantum state. We shall refer to this as the classical limit of the Wigner function. Furthermore, in this sense, we can identify our action variables with a constant.

Using this formalism it was shown in  \cite{DGGV2022} that in the classical limit (which we denote by $\simeq$)
\begin{equation}
 \boldsymbol{\sigma} \simeq \boldsymbol{\sigma}^{\mathrm{cl}}\, , \label{eq:equicovma}
\end{equation}
where $\boldsymbol{\sigma}^{\mathrm{cl}}= (\sigma^{\mathrm{cl}}_{\alpha \beta})$ is the classical covariance matrix with elements given by
\begin{equation}
\sigma^{\mathrm{cl}}_{\alpha \beta} =  \langle r_{\alpha} r_{\beta} \rangle_{\mathrm{cl}}-\langle r_{\alpha} \rangle_{\mathrm{cl}} \langle  r_{\beta} \rangle_{\mathrm{cl}} \,. \label{eq:sigmac}
\end{equation}
Here, $r=\{r_{\alpha}\}= (q_1,\dots,q_N,$ $p_1,\dots,p_N)^{\intercal}$ is a phase-space column vector and
\begin{equation}\label{classAvg}
\langle f \rangle_{\mathrm{cl}}= \frac{1}{\left( 2 \pi\right)^N}  \int_{0}^{2 \pi} \!\! \cdots  \int_{0}^{2 \pi} \!\!\! \mathrm{d}^N \varphi \, f  \,,
\end{equation}
with $\mathrm{d}^N \varphi= {\rm d} \varphi_1 \, \dots \, {\rm d} \varphi_N$, is the classical average of a function $f=f(\varphi,I)$ over the angle variables.

The relation \eqref{eq:equicovma} was used in \cite{DGGV2022} to define classical analogs of some quantum quantities that serve as a measure of the mixedness of the quantum states and that for Gaussian states can be written as functions depending only on the quantum covariance matrix. Such quantities are the purity $\mu$, linear quantum entropy $S_L$, and  von Neumann entropy $S$ which, for a normalized quantum state described by a density operator $\hat{\boldsymbol{\rho}}$,  are defined as~\cite{deGosson2006, diosi2011}
\begin{subequations} \label{eq:qq}
\begin{align}
\mu \left( \hat{\boldsymbol{\rho}}\right)&= \Tr \hat{\boldsymbol{\rho}}^2\,,   \label{eq:pur}\\
S_{\mathrm{L}} \left( \hat{\boldsymbol{\rho}}\right) &= 1-\mu\left( \hat{\boldsymbol{\rho}}\right)\,, \label{eq:lS}\\
S \left( \hat{\boldsymbol{\rho}}\right)&= -\Tr \left( \hat{\boldsymbol{\rho}}  \ln \hat{\boldsymbol{\rho}}  \right)\,. \label{eq:entro}
\end{align}
\end{subequations}
For pure states $\hat{\boldsymbol{\rho}}^2 = \hat{\boldsymbol{\rho}}$, $\mu$ takes a maximum value of $1$ and both entropies $S_{\mathrm{L}}$ and $S$ are zero, whereas for mixed states, $\hat{\boldsymbol{\rho}}^2 \neq \hat{\boldsymbol{\rho}}$, these  quantities vary in the range of $0< \mu < 1$, $ 0< S_{\mathrm{L}} < 1$ and $ S >0$.

For a Gaussian state, the quantities \eqref{eq:qq} depend only on the corresponding quantum covariance matrix \cite{Agarwal1971, Holevo1999, Dodonov_2002, Paris2003, deGosson2006, diosi2011, Golubeva2014, serafini2017, deGosson2019, Demarie2018}. In fact, considering an $n$-mode Gaussian state ($n$ denotes the degrees of freedom of the subsystem formed by the particles $a_1,\dots,a_n$ of the system of $N$ degrees of freedom) with the reduced quantum covariance matrix $\boldsymbol{\sigma}_{(n)}$, we have
 \begin{subequations} \label{eq:qqf}
 \begin{align}
\mu \left(a_1,a_2,\dots,a_n\right)&= \left( \frac{\hbar}{2}\right)^n \frac{1}{\sqrt{\det \boldsymbol{\sigma}_{(n)} }} \label{eq:qpu}\\
&= \frac{1}{2^n}  \prod_{k=1}^n \nu_k^{-1}\,, \nonumber \\
S_{\mathrm{L}}\left( a_1,a_2,\dots,a_n\right)&=1- \left( \frac{\hbar}{2}\right)^n \frac{1}{\sqrt{\det \boldsymbol{\sigma}_{(n)} }} \,, \label{eq:qle}\\
    S \left( a_1,a_2,\dots,a_n\right)&= \sum_{k=1}^{n} \mathcal{S} (\nu_k)\,, \label{eq:Sred}
 \end{align}
\end{subequations}
where
\begin{equation}
   \mathcal{S}(\nu_k)=  \left( \nu_k+\frac{1}{2} \right)\ln \left( \nu_k+\frac{1}{2} \right)-\left( \nu_k-\frac{1}{2} \right) \ln \left( \nu_k-\frac{1}{2} \right)\,, 
\end{equation}
the subscript $(n)$ represents $(a_1,a_2,\dots,a_n)$, and $\nu_k$ are the symplectic eigenvalues of $\boldsymbol{\sigma}_{(n)} / \hbar$, i.e., they are the entries of a nonnegative diagonal matrix $\boldsymbol{D}=\textrm{diag}\{ \nu_1,\dots,\nu_n\}$ which, together with a suitable symplectic matrix $\boldsymbol{M}$, allows us to write
\begin{align}
    \boldsymbol{M}^{\intercal}\left(\frac{\boldsymbol{\sigma}_{(n)}}{\hbar} \right) \boldsymbol{M}&= \left( \begin{array}{ll}
\boldsymbol{D}         &  \mathbf{0}_{n \times n}\\
\mathbf{0}_{n \times n}        &  \boldsymbol{D}
    \end{array}\right)\,. \label{symp}
\end{align}
Notice that $\mathcal{S}(\nu_k)=0$ only if $\nu_k=1/2$, and that in the case $n=1$, for the particle $a_i$,  we have \cite{Holevo1999}
\begin{align}
\nu_1=  \frac{1}{\hbar} \sqrt{   \sigma_{{\bf{p}}_{a_i} {\bf{p}}_{a_i} } \sigma_{{\bf{q}}_{a_i} {\bf{q}}_{a_i} } - \left(\sigma_{{\bf{q}}_{a_i} {\bf{p}}_{a_i} }\right)^2}\,.
\end{align}

Bearing in mind  the relation \eqref{eq:equicovma} $(\boldsymbol{\sigma}_{(n)}\simeq \boldsymbol{\sigma}_{(n)}^{\mathrm{cl}})$, the quantum functions \eqref{eq:qqf}, and the Bohr-Sommerfeld quantization rule for the action variables, in the sense $ \hbar/2 \to I_k $, the {\it classical analogs} of the purity \eqref{eq:qpu}, linear quantum entropy \eqref{eq:qle}, and von Neumann entropy \eqref{eq:Sred} are, respectively, defined as~\cite{DGGV2022}

\begin{subequations}\label{eq:cq1}
\begin{align}
\tilde{\mu}^{\mathrm{cl}} (a_1,a_2,\dots,a_n)&:= c_1^{n}  \lim_{I_a \to c_1} \frac{1}{\sqrt{\det \boldsymbol{\sigma}_{(n)}^{\mathrm{cl}}}}   \label{eq:pca}\\
&=\left( \frac{1}{2^n}\right)  \prod_{k=1}^n \tilde{\sigma}_k^{-1} \,, \nonumber\\
\tilde{S}^{\textrm{cl}}_L(a_1,a_2,\dots,a_n)&:= 1-c_1^{n}  \lim_{I_a \to c_1} \frac{1}{\sqrt{\det \boldsymbol{\sigma}_{(n)}^{\mathrm{cl}}}} \,,\label{eq:leal} \\
\tilde{S}^{\textrm{cl}} \left( a_1,a_2,\dots,a_n \right) &:=  \sum_{k=1}^{n} \mathcal{S}^{\textrm{cl}}(\tilde{\sigma}_k) \,, \label{eq:caen}
\end{align}
\end{subequations}
where
\begin{subequations}
\begin{align}
    \mathcal{S}^{\textrm{cl}}(\tilde{\sigma}_k):=&  \left( \tilde{\sigma}_k+\frac{1}{2} \right)\ln \left( \tilde{\sigma}_k+\frac{1}{2} \right) \nonumber\\
    &-\left( \tilde{\sigma}_k-\frac{1}{2} \right) \ln \left(\tilde{\sigma}_{k}-\frac{1}{2} \right)\,, \label{indientro}\\
     \tilde{\sigma}_k :=&   \frac{1}{2 c_2}  \lim_{I_a \to c_2} \sigma^{\textrm{cl}}_{k} \,, \label{eq:cleige}
\end{align}
\end{subequations}
with $\sigma^{\textrm{cl}}_{k}$ the symplectic eigenvalues of $\boldsymbol{\sigma}^{\textrm{cl}}_{(n)}$, and $I_{a_k}$ the action variables associated with the $k$-th normal mode. Note that $\tilde{\sigma}_k$ are the symplectic eigenvalues of $1/(2c_2) \lim_{I_k \to c_2} \boldsymbol{\sigma}{(n)}^{\mathrm{cl}}$ (analogous to $\nu_k$ which are the symplectic eigenvalues of $\boldsymbol{\sigma}{(n)}/\hbar$). It's important to note that $c_1$ and $c_2$ are real positive constants that disappear during the calculation of quadratic Hamiltonian systems, which in the quantum context give rise to Gaussian states in their ground state. This is because, for these systems, the $q$ and $p$ variables depend linearly on the square root of $I_k$ [see (17)]. Then, by identifying all $I_k$ as $c_2$, the classical covariance matrix becomes proportional to $c_2$ [see (19)], causing $\tilde{\sigma}_k$ to be independent of $c_2$.

In the particular case of $n=1$, for the particle $a_i$ we get
\begin{equation}
  \tilde{\sigma}_1 =  \frac{1}{2c_2} \lim_{I_{a} \to c_2}   \sqrt{ \sigma^{\textrm{cl}}_{p_{a_i} p_{a_i} } \sigma^{\textrm{cl}}_{q_{a_i} q_{a_i} } - \left(\sigma^{\textrm{cl}}_{q_{a_i} p_{a_i} }\right)^2}\,.
\end{equation}
Notice that  the limits in \eqref{eq:pca} and \eqref{eq:cleige} involve not only a particular action variable $I_{a_k}$, but all the action variables.

We must note that to calculate the classical functions \eqref{eq:cq1} no a priori knowledge of the corresponding quantum system is required. In addition, these functions are defined for any integrable system. However, it is only when the corresponding quantum system is in a Gaussian state that \eqref{eq:cq1} are the classical analogs of the quantum purity, linear quantum entropy, and von Neumann entropy, respectively.

\subsection*{Example: Generalized harmonic oscillator chain}\label{GHOC}

Let us now illustrate how to compute the classical analog of the purity. To this end let us take a generalized harmonic oscillator chain (GHOC) consisting of $N$ coupled oscillators \cite{chruscinski}. The Hamiltonian of the system is
\begin{align}
\label{hamiltonianoqp}
H(q,p)&=\frac{1}{2} p^{\intercal} p+\frac{1}{2} q^{\intercal} \boldsymbol{K} q+q^{\intercal} \boldsymbol{Y} p\,, 
\end{align}
where $p=(p_1,\dots,p_N)^{\intercal}$, $q=(q_1,\dots,q_N)^{\intercal}$, $\boldsymbol{Y}$ is an $N\times N$ diagonal matrix whose diagonal elements are parameters denoted by $Y_a$, and  $\boldsymbol{K}$ is $N\times N$ symmetric matrix of parameters.

Before starting we would like to make some general assertions about the stability of the system. We first determine the fixed points. The equations of motion coming from  \eqref{hamiltonianoqp} are
\begin{align}
\label{hamiltonequations}
\left(\begin{array}{c} \Dot{q}  \\ \Dot{p}\end{array}\right)=\boldsymbol{A}\left(\begin{array}{c} q  \\ p \end{array}\right)\, ,
\end{align}
where
\begin{align}
\boldsymbol{A}=\left(\begin{array}{cc}  \boldsymbol{Y} \quad& \boldsymbol{1}_{N \times N} \\-\boldsymbol{K} \quad &   \boldsymbol{Y}
\end{array}\right)\,.
\end{align}
Setting $(\Dot{q},\Dot{p})=(0,0)^{\intercal}$, it is direct to see that there is a unique fixed point  at $(q^*,p^*)^{\intercal}= (0,0)^{\intercal}$. To analyze the stability of this equilibrium point, we calculate the eigenvalues $\lambda$ of the matrix $\boldsymbol{A}$. The characteristic equation for $\lambda$ is $\det (\boldsymbol{M}+\lambda^2\boldsymbol{1})=0$ where $ \boldsymbol{M}:= \boldsymbol{K}-\boldsymbol{Y}^2$. Then, denoting by $\omega_a^2$ the eigenvalues of $\boldsymbol{M}$, the eigenvalues of the $\boldsymbol{A}$ are $ \lambda_a^{(\pm)}=\pm i \omega_a$. If we restrict the system to $\boldsymbol{M}>0$,  then $\omega_a$ are real and $\lambda_a^{(\pm)}$ are purely imaginary. This corresponds to a phase portrait of a center which is a stable system. On the other hand,  if the matrix $\boldsymbol{M}$ is not positive semidefinite, we have at least one imaginary $\omega_a$, and hence $\boldsymbol{A}$ has at least one positive and one negative eigenvalue, which originates an unstable hyperbolic system. Therefore, the condition $\boldsymbol{M}>0$ guarantees that the system is stable. In the following we consider $\boldsymbol{M}>0$.

To compute the classical analog of the purity, we start by performing the canonical transformation $q=q^{\prime}$, $p=p^{\prime}-\boldsymbol{Y}q$, followed by the canonical transformation of the form $Q=\boldsymbol{S}^{\intercal} q^{\prime}$, $ P=\boldsymbol{S}^{\intercal} p^{\prime}$, where $\boldsymbol{S}$ is such that it diagonalizes the matrix $ \boldsymbol{M}$, i.e., $\boldsymbol{S}\boldsymbol{M}\boldsymbol{S}^{\intercal}=\boldsymbol{\Omega}^2$ where $\boldsymbol{\Omega}={\rm diag}(\omega_1,\dots,\omega_N)$. Thus, the total canonical transformation can be written as $ q=\boldsymbol{S} Q$, $p=\boldsymbol{S}P - \boldsymbol{Y} \boldsymbol{S}Q$. In terms of the new variables $(Q,P)$ the Hamiltonian \eqref{hamiltonianoqp} reads
\begin{align} \label{hamiltonianoqp2}
H (Q, P) & =\frac{1}{2} P^{\intercal} P+\frac{1}{2} Q^{\intercal} \boldsymbol{\Omega}^2 Q \,. 
\end{align}
From \eqref{hamiltonianoqp2} it is clear that $\omega_a$ are the normal frequencies of the system.

In terms of action-angle variables $(I, \phi)$, the new variables $(Q,P)$ can be written as $Q_a=\sqrt{\tfrac{2I_a}{\omega_a}} \sin\phi_a$,
$P_a=\sqrt{2 I_a \omega_a} \sin\phi_a$, and therefore
\begin{subequations}\label{qpactionangle}
\begin{align}
q_a& =\sum_b S_{ab}\sqrt{\dfrac{2I_b}{\omega_b}}\sin\phi_b\,,\\
p_a&=\sum_b \left(S_{ab}\sqrt{2I_b \omega_b}\cos\phi_b-Y_a S_{ab}\sqrt{\dfrac{2I_b}{\omega_b}}\sin\phi_b\right)\,.
\end{align}
\end{subequations}

Using (\ref{classAvg}) and \eqref{qpactionangle}, the terms involved in the entries of the classical covariance matrix \eqref{eq:sigmac} are
\begin{subequations}\label{averCV}
\begin{align}
\langle q_a q_b \rangle_{\mathrm{cl}}&= \left(\boldsymbol{S} \boldsymbol{I}\boldsymbol{\Omega}^{-1}\boldsymbol{S}^{\intercal}  \right)_{ab}\,,\\
\langle q_a p_b \rangle_{\mathrm{cl}}&= \left(\boldsymbol{S} \boldsymbol{I}\boldsymbol{\Omega}^{-1}\boldsymbol{S}^{\intercal}\boldsymbol{Y}  \right)_{ab}\,,\\
\langle p_a p_b \rangle_{\mathrm{cl}}&= \left(\boldsymbol{S}\boldsymbol{I}\boldsymbol{\Omega}\boldsymbol{ S}^{\intercal} + \boldsymbol{Y}\boldsymbol{S} \boldsymbol{I}\boldsymbol{\Omega}^{-1}\boldsymbol{S}^{\intercal}\boldsymbol{Y}  \right)_{ab}\,, \label{18c}\\
\langle q_a \rangle_{\mathrm{cl}}&=  \langle p_a \rangle_{\mathrm{cl}}=0\,,
\end{align} 
\end{subequations}
where $\boldsymbol{I}={\rm diag}(I_1,\dots,I_N)$. Notice that as a consequence of the term $q^{\intercal} \boldsymbol{Y} p$ in the Hamiltonian \eqref{hamiltonianoqp}, we have $\langle q_a p_b \rangle_{\mathrm{cl}}\neq 0$ and the additional term $\boldsymbol{Y}\boldsymbol{S} \boldsymbol{I}\boldsymbol{\Omega}^{-1}\boldsymbol{S}^{\intercal}\boldsymbol{Y}$ in \eqref{18c}. Considering that all action variables are equal to some constant $c_1$, i.e., $I_a \to c_1$ for all $a=1,\dots,N$, the classical covariance matrix takes the form
\begin{align} \label{covsimpex2}
\lim_{I_a \to c_1}  \boldsymbol{\sigma}^{\mathrm{cl}} =c_1\left(\begin{array}{cc} \boldsymbol{M}^{-1/2} & -\boldsymbol{M}^{-1/2} \boldsymbol{Y} \\-\boldsymbol{Y} \boldsymbol{M}^{-1/2}\;\; &  \boldsymbol{M}^{ 1/2}+\boldsymbol{Y} \boldsymbol{M}^{-1/2}\boldsymbol{Y}\end{array}\right)\,,
\end{align}
where $\boldsymbol{M}^{-1/2}$ is the inverse of the matrix $\boldsymbol{M}^{1/2}= \boldsymbol{S} \boldsymbol{\Omega} \boldsymbol{S}^{\intercal}$.

Using Schur complement is no hard to show that the determinant of \eqref{covsimpex2} is $c_1^{2N}$, and then according to  \eqref{eq:pca}, the classical analog of purity of the whole system is 1, which is the same result that follows from the quantum frame. In general, to compute the purity of a subsystem consisting of the first $m$ ($m \leq N$) oscillators of the system, we need the associated reduced covariance matrix, which is obtained by taking the corresponding $m\times m$ block-submatrix from \eqref{covsimpex2}. The reduced covariance matrix $\boldsymbol{\sigma}_{(m)}^{\mathrm{cl}}$ can be written as

\begin{align}\label{redcovex2}
\boldsymbol{\sigma}_{(m)}^{\mathrm{cl}}=c_1\left(\begin{array}{cc} \boldsymbol{D} & -\boldsymbol{D} \boldsymbol{Y}_{red} \\- \boldsymbol{Y}_{red}\boldsymbol{D} & \boldsymbol{A}+\boldsymbol{Y}_{red} \boldsymbol{D}\boldsymbol{Y}_{red}\end{array}\right)\,,
\end{align}
where $\boldsymbol{A}$ and  $\boldsymbol{D}$  are block matrices of  $\boldsymbol{M}^{1/2}$ and $\boldsymbol{M}^{-1/2}$, respectively,
\begin{subequations}
\begin{align}
\boldsymbol{M}^{1/2}&=\left(\begin{array}{cc} \boldsymbol{A} & \boldsymbol{B} \\ \boldsymbol{B}^{\intercal} & \boldsymbol{C}\end{array}\right)\,,\\
\boldsymbol{M}^{-1/2}&=\left(\begin{array}{cc} \boldsymbol{D} & \boldsymbol{E} \\ \boldsymbol{E}^{\intercal} & \boldsymbol{F}\end{array}\right)\,,
\end{align}
\end{subequations}
and $\boldsymbol{Y}_{red}=\operatorname{diag}(Y_1,\dots,Y_m)$.

Using \eqref{eq:pca} and \eqref{redcovex2}, the classical analog of purity of the $m$ oscillators turns out to be 
\begin{align}\label{classpurity}
\tilde{\mu}^{\mathrm{cl}}(a_1,\dots,a_m)=\dfrac{1}{\sqrt{\det \boldsymbol{A} \det \boldsymbol{D} }}\, ,
\end{align}
where we have used Schur complement to simplify the determinant. Notice that even though the matrix $\boldsymbol{Y}_{red}$ does not appear explicitly in this result, the classical analog of the purity depends on the parameters $Y_a$ through the frequencies $\omega_a$ which are involved in $\boldsymbol{A}$ and $\boldsymbol{D}$. We show below that this result is exactly the predicted one by the quantum purity \eqref{eq:qpu}.

Before going to the quantum framework, we consider a particular case of \eqref{hamiltonianoqp} and provide an explicit expression for the classical analog of the purity. Let us take the system of two coupled generalized harmonic oscillators described by the Hamiltonian
\begin{align}\label{Hcasoparticular}
H(q,p)=H_{1}+H_{2}+\frac{1}{2} Z\left(q_{1}-q_{2}\right)^{2}\,,
\end{align}
where
\begin{align}
H_{a} (q_a,p_a)=\frac{1}{2}\left(p_{a}^{2}+Y_a\left(q_{a} p_{a}+p_{a} q_{a}\right)+X_a q_{a}^{2}\right)\,.
\end{align}
In this case, the matrices $\boldsymbol{K}$ and $\boldsymbol{M}$ are 
\begin{subequations}
\begin{align}
\boldsymbol{K}=&\left(\begin{array}{cc} X_1+Z& -Z \\-Z & X_2+Z\end{array}\right)\,,\\  \boldsymbol{M}=&\left(\begin{array}{cc}X_1+Z-Y_1^{2} & -Z \\-Z & X_2+Z-Y_2^{2}\end{array}\right) \,,
\end{align}
\end{subequations}
and hence the matrix $\boldsymbol{S}$ is
\begin{align}
\boldsymbol{S}=\left(\begin{array}{cc}\cos \theta & -\sin\theta \\\sin\theta & \cos\theta \end{array}\right)\,, 
\end{align}
where $\tan\theta=\tfrac{\gamma}{|\gamma|}\left( \gamma^2+1 \right)-\gamma$ with $\gamma:=\tfrac{X_2-X_1+Y_1^2-Y_2^2}{2Z}$. Here, we have considered $X_2-X_1+Y_1^2-Y_2^2\neq0$ and $Z\neq 0$. Moreover, the normal frequencies are 
\begin{subequations} \label{eq:frecuenciasgen}
\begin{align}
\omega_1&=\sqrt{X_1-Y_1^2+Z-Z\tan\theta}\,,\\
\omega_2&=\sqrt{X_2-Y_2^2+Z+Z\tan\theta}\,. 
\end{align}
\end{subequations}

Having these matrices at hand and using \eqref{averCV}, we can compute the classical covariance matrix \eqref{covsimpex2}. The result is
\begin{align} \label{classCVcase1}
\lim_{I_a \to c_1}  \boldsymbol{\sigma}^{\mathrm{cl}} =\left(\begin{array}{cc} \boldsymbol{\sigma}_{qq}^{\mathrm{cl}} & \boldsymbol{\sigma}_{qp}^{\mathrm{cl}} \\\boldsymbol{\sigma}_{pq}^{\mathrm{cl}}\;\; &  \boldsymbol{\sigma}_{pp}^{\mathrm{cl}}\end{array}\right)\,,
\end{align}
where
\begin{widetext}
\begin{subequations}\label{compccmex2}
\begin{align}
\boldsymbol{\sigma}_{qq}^{\mathrm{cl}}&=c_1 \left(\begin{array}{cc}\frac{\cos^2\theta}{\omega_1}+\frac{\sin^2\theta}{\omega_2} & \left(\frac{1}{\omega_{1}}-\tfrac{1}{\omega_2}\right)\sin\theta\cos\theta \\ \left(\frac{1}{\omega_{1}}-\frac{1}{\omega_2}\right)\sin\alpha\cos\theta  & \tfrac{\sin^2\theta}{\omega_1}+\frac{\cos^2\theta}{\omega_2} \end{array}\right) \,,\\
 \boldsymbol{\sigma}_{q p}^{\mathrm{cl}}&=\boldsymbol{\sigma}_{p q}^{\mathrm{cl}} {}^{\intercal}=c_1 \left(\begin{array}{cc}Y_1\left(\frac{\cos^2\theta}{\omega_1}+\frac{\sin^2\theta}{\omega_2}\right) & Y_2\left(\frac{1}{\omega_{1}}-\tfrac{1}{\omega_2}\right)\sin\theta\cos\theta  \\ Y_1\left(\frac{1}{\omega_{1}}-\frac{1}{\omega_2}\right)\sin\theta\cos\theta   & Y_2\left(\tfrac{\sin^2\theta}{\omega_1}+\frac{\cos^2\theta}{\omega_2}\right) \end{array}\right) \,, \\
\boldsymbol{\sigma}_{pp}^{\mathrm{cl}}&=c_1\left(\begin{array}{cc}\omega_1\cos^2\theta + \omega_2\sin^2\theta +  Y_1^2\left(\frac{\cos^2\theta}{ \omega_{1}}+\frac{\sin^2\theta}{\omega_2}\right) \;\; &\left(\left(\omega_1 - \omega_2\right) +  Y_1Y_2\left(\frac{1}{ \omega_{1}}-\frac{1}{\omega_2}\right)\right)\sin\theta\cos\theta \\ \left(\left(\omega_1 - \omega_2\right) +  Y_1Y_2\left(\frac{1}{ \omega_{1}}-\frac{1}{\omega_2}\right)\right)\sin\theta\cos\theta \;\;&  \omega_1\sin^2\theta + \omega_2\cos^2\theta +  Y_2^2\left(\frac{\sin^2\theta}{ \omega_{1}}+\frac{\cos^2\theta}{\omega_2}\right)  \end{array}\right)\,.
\end{align}
\end{subequations}

Using \eqref{classpurity} together with \eqref{classCVcase1}, we find the classical analog of purity for a subsystem of one particle
\begin{align}\label{capex2}
\tilde{\mu}^{\mathrm{cl}}(a_1)= \sqrt{\frac{\omega_1\omega_2}{\left( \omega_1\cos^2\theta+\omega_2\sin^2\theta \right) \left( \omega_2\cos^2\theta+\omega_1\sin^2\theta \right)}}\,.
\end{align}
\end{widetext}

Notice that if  $\omega_1= \omega_2$, then \eqref{capex2} reduces to $\tilde{\mu}(a_1)= 1$. Nevertheless, in this case, $\gamma$ is imaginary because of the condition $Z \neq 0$, i.e., since the oscillators are coupled. That $\gamma$ is imaginary implies that the corresponding quantum Hamiltonian is non-Hermitian. The quantization of this type of systems has been done using PT-symmetry \cite{Bender, Moiseyev}, and it has been discovered that entanglement has different properties in this context \cite{PTsymmetry,Non-Hermitian}. In the general case,  $\omega_1 \neq \omega_2$, from \eqref{capex2} it follows that $\tilde{\mu}(a_1) < 1$ and hence the quantum counterpart of the system is entangled.

We now turn to the quantum framework. The quantum counterpart of the Hamiltonian (\ref{hamiltonianoqp}) is
\begin{align}\label{qHamil}
\hat{\bf{H}} (\hat{\bf{q}}, \hat{\bf{p}}) &=\frac{1}{2} \hat{\bf{p}}^{\intercal} \hat{\bf{p}}+\frac{1}{2} \hat{\bf{q}}^{\intercal} \boldsymbol{K} \hat{\bf{q}}+\frac{1}{2}\left( \hat{\bf{q}}^{\intercal} \boldsymbol{Y} \hat{\bf{p}}+\hat{\bf{p}}^{\intercal} \boldsymbol{Y} \hat{\bf{q}}\right)\,. 
\end{align}
Performing a transformation analogous to the one that leads to \eqref{hamiltonianoqp2}, the Hamiltonian \eqref{qHamil} can be diagonalized as $\hat{\bf{H}}(\hat{\bf{Q}},\hat{\bf{P}}) =\frac{1}{2} \hat{\bf{P}}^{\intercal} \hat{\bf{P}}+\frac{1}{2} \hat{\bf{Q}}^{\intercal} \boldsymbol{\Omega}^2 \hat{\bf{Q}}$. To compute the components \eqref{qumet} of the quantum covariance matrix for the ground (Gaussian) state, it is convenient to write the operators $\hat{\bf{Q}}$ and $\hat{\bf{P}}$ as a combination of the usual raising and lowering operators. By doing this, the resulting quantum covariance matrix is the same as the classical covariance matrix~\eqref{covsimpex2}, but with $\hbar/2$ replacing the constant $c_1$. Using this result together with \eqref{eq:qpu}, the quantum purity of the $m$ quantum oscillators turns out to be equal to \eqref{classpurity}. This illustrates that the classical quantity $\tilde{\mu}^{\mathrm{cl}}$ is capable of providing the same mathematical results as its quantum counterpart $\mu$.

\section{Generalized purities and entropies, and logarithmic negativity}\label{sec3}

\subsection{Generalized purities and entropies}

 The quantum quantities \eqref{eq:qq} can be generalized in the following sense: for a given quantum state $\hat{\boldsymbol{\rho}}$, the generalized purities, Bastiaans-Tsallis entropies, and R\'enyi entropies \cite{renyi1970} for $\alpha \geq 0$ are given  by \cite{serafini2017,Bastiaans84, Tsallis88, adesso2014}
\begin{subequations}\label{genequa}
 \begin{align}
\mu_{\alpha} \left( \hat{\boldsymbol{\rho}} \right) &= \Tr  \hat{\boldsymbol{\rho}}^{\alpha}\,, \label{genpur}\\
 S_{\alpha} \left( \hat{\boldsymbol{\rho}} \right) &= \frac{1-\Tr  \hat{\boldsymbol{\rho}}^{\alpha}}{\alpha-1}  \,, \label{genlien}\\
 H_{\alpha} \left(  \hat{\boldsymbol{\rho}} \right) &= \frac{ \ln \left( \Tr  \hat{\boldsymbol{\rho}}^{\alpha} \right) }{1-\alpha}\,, \label{genrenen}
 \end{align}
\end{subequations}
respectively. Some comments regarding these generalizations. First, notice that $ \Tr  \hat{\boldsymbol{\rho}}^{\alpha}= \left( ||  \hat{\boldsymbol{\rho}} ||_{\alpha} \right)^{\alpha}$ where $|| \,\, ||_{\alpha}$ is the Schatten $\alpha$-norm \cite{Rajendra, serafini2017} and, in the asymptotic limit of arbitrary large $\alpha$, $\Tr \hat{\boldsymbol{\rho}}^{\alpha}$ is a function of the largest eigenvalue of $\hat{\boldsymbol{\rho}}$ only. Second, \eqref{genpur} reduces to the purity (\ref{eq:pur}) when $\alpha=2$. Third, for $\alpha=2$, \eqref{genlien} yields the linear entropy \eqref{eq:lS}, while in the limit $\alpha \to 1$ it reduces to the von Neumann entropy \eqref{eq:entro}. Fourth, in the limit $\alpha \to 1$, \eqref{genrenen} becomes the von Neumann entropy \eqref{eq:entro}.  Fifth, in the limit $\alpha \to \infty$, $S_\alpha$ goes to a trivial constant null function, losing all information about the quantum state, while $H_{\alpha}$ converges to the min-entropy, which is the smallest entropy measure in the family of R\'enyi entropies \cite{WEHRL1991, Linden2013}. Sixth, $S_\alpha$ is a monotonically decreasing function of $\alpha$, for a given quantum state.

For a $n$-mode Gaussian state, associated with a subsystem formed by the particles $a_1, \dots, a_n$,  \eqref{genequa} can be written as functions that only depend on the covariance matrix $\boldsymbol{\sigma}_{(n)}$. In fact, they are given by  \cite{Adesso2004,adesso2014, Kim2018}
\begin{subequations}\label{genequagaussian}
\begin{align}
\mu_{\alpha} \left(a_1,a_2,\dots,a_n\right) & =\prod_{k=1}^{n} g_{\alpha} (\nu_k)\,,\\
     S_{\alpha} \left(a_1,a_2,\dots,a_n\right) &=  \frac{1- \prod_{k=1}^{n} g_{\alpha} (\nu_k)  }{\alpha-1} \,,\\
      H_\alpha \left(a_1,a_2,\dots,a_n\right)&= \frac{\sum_{k=1}^{n} \ln \left( g_{\alpha} (\nu_k) \right)}{1-\alpha} \,,
\end{align}
\end{subequations}
where $\nu_k$ are the symplectic eigenvalues of $\boldsymbol{\sigma}_{(n)} / \hbar$, and
\begin{equation}
    g_\alpha(\nu_k)= \frac{1}{ \left( \nu_k+\frac{1}{2}\right)^\alpha -\left( \nu_k-\frac{1}{2}\right)^\alpha} \,. \label{functiong}
\end{equation}

Notice that $g_\alpha(1/2)=1$ only if all eigenvalues satisfy $\nu_k=1/2$, which implies $\mu_{\alpha}=1,S_{\alpha}=0,$ and $H_\alpha=0$.

\subsection{Logarithmic negativity}

We now focus on the logarithmic negativity for a Gaussian state in a system of $N$  coupled harmonic oscillators. We begin by considering a set of $m=n_{1}+n_{2}\leqslant N$ oscillators that are divided into two groups; one of them with $n_{1}$ oscillators and another with $n_{2}$ oscillators. Assuming that there are no correlations between positions and momenta, the quantum covariant matrix associated with the $m$ oscillators, which we refer to as the reduced covariance matrix $\boldsymbol{\mu}$, can be obtained from the quantum covariance matrix $\boldsymbol{\sigma}$ of the entire system by taking the rows and columns corresponding to the $m$ oscillators. Then, under this condition, the matrix $\boldsymbol{\mu}$ has the form
\begin{equation}
  \boldsymbol{\mu}=\frac{1}{2}\left(\begin{array}{cc} \boldsymbol{\mu}_{q} & \boldsymbol{0}_{m \times m} \\ \boldsymbol{0}_{m \times m} & \boldsymbol{\mu}_{p} \end{array}\right),
\end{equation}
where $\boldsymbol{\mu}_{q}$ and $\boldsymbol{\mu}_{p}$ are $m \times m$ matrices. Let us now consider the partial transpose $\boldsymbol{\mu}^{\Gamma}$ of $\boldsymbol{\mu}$ with respect to the group of $n_{2}$ oscillators, which is defined as 
\begin{equation}
 \boldsymbol{\mu}^{\Gamma}:=\boldsymbol{P} \boldsymbol{\mu} \boldsymbol{P}, \label{covred}
\end{equation}
where $\boldsymbol{P}$ is a diagonal matrix given by
\begin{equation}
 \boldsymbol{P}=\left(\begin{array}{cc} \boldsymbol{1}_{m \times m} & \boldsymbol{0}_{m \times m} \\ \boldsymbol{0}_{m \times m} & \boldsymbol{P}_{p}\end{array}\right), \label{matrixP}
\end{equation}
with $\boldsymbol{P}_{p}= \operatorname{diag}(1, \dots, 1,-1, \dots,-1)$. The entries of $\boldsymbol{P}_{p}$ are $1$ for an oscillator of the group with $n_{1}$ oscillators or $-1$ for an oscillator belonging to the group with $n_{2}$ oscillators. It is not hard to verify that the effect of partial transposition with respect to the group of $n_{2}$ oscillators is to change the sign of the momenta corresponding to these oscillators.

To compute the logarithmic negativity we also require the symplectic matrix $\boldsymbol{\Omega}=(\Omega_{\alpha \beta})$, whose entries are given by 
\begin{equation}
\Omega_{\alpha \beta}:=-i\left[\hat{\bf{r}}_{\alpha}, \hat{\bf{r}}_{\beta}\right],   
\end{equation}
and has the block matrix form
\begin{equation}
\boldsymbol{\Omega}=\hbar\left(\begin{array}{ll}
\boldsymbol{0}_{N \times N} & \boldsymbol{1}_{N \times N} \\
-\boldsymbol{1}_{N \times N} & \boldsymbol{0}_{N \times N}
\end{array}\right).
\end{equation}
Thus, for the set of $m$ oscillators the associated symplectic matrix $\boldsymbol{\Sigma}$ takes the form
\begin{equation}
\boldsymbol{\Sigma}:=\hbar\left(\begin{array}{ll}
\boldsymbol{0}_{m \times m} & \boldsymbol{1}_{m \times m} \\
-\boldsymbol{1}_{m \times m} & \boldsymbol{0}_{m \times m}
\end{array}\right).\label{sympred}
\end{equation}

Using \eqref{covred}, \eqref{sympred}, and considering the ground state of the chain, the logarithmic negativity $E_{\mathcal{N}}$, which provides a measure of entanglement between the two groups of $n_1$  and $n_2$ oscillators, is given by~\cite{Vidal2002}
\begin{equation}
    E_{\mathcal{N}}=-\sum_{k=1}^{2 m} \log _{2}\left[\min \left(1, \mid \lambda_{k} \mid \right)\right] \,, \label{LogNeg}
\end{equation}
where $\lambda_{k}$ ($k=1, \dots, 2 m)$ are the eigenvalues of the matrix
\begin{equation}
    \boldsymbol{B}={\rm i} \boldsymbol{\Sigma}^{-1} \boldsymbol{\mu}^{\Gamma}. \label{Bquantum}
\end{equation}
In fact, if $E_{\mathcal{N}}$ is positive, then the two groups of $n_1$ and $n_2$ oscillators are entangled. In terms of the matrices $\boldsymbol{\mu}_{q}$ and $\boldsymbol{\mu}_{p}$, the logarithmic negativity can be written as \cite{Audenaert2002}
\begin{equation}
  E_{\mathcal{N}} =- \sum_{j=1}^{m} \log _{2}\left[\min \left(1, \tilde{\lambda}_{j}\right)\right] \label{LogNeg2},
\end{equation}
where $\tilde{\lambda}_{j}$ are the eigenvalues of the matrix $\boldsymbol{\mu}_{q} \boldsymbol{P}_{p} \boldsymbol{\mu}_{p} \boldsymbol{P}_{p}/\hbar^{2}$.

Before concluding this section, it is worth pointing out that another useful measure of entanglement in composite systems is the {\it negativity} $\mathcal{N}$~\cite{Vidal2002}, which is related to the logarithmic negativity as
\begin{equation}
    \mathcal{N}=\frac{2^{E_{\mathcal{N}}}-1}{2}. \label{Negativity}
\end{equation}

\section{Classical analogs} \label{sec4}

\subsection{Generalized purities and entropies}\label{subsecGPE}

In this subsection, we define classical analogs of the generalized purities and entropies for Gaussian states \eqref{genequagaussian}, in the framework of classical integrable systems.  We consider here a subsystem consisting of the $n$ particles $a_1,a_2,\dots,a_n$ of a classical integrable system of $N$ degrees of freedom, which has been written in terms of the action-angle variables $(\varphi, I)$.

We start by introducing the classical analog of the function $g_\alpha$ for the subsystem ($a_1,a_2,\dots,a_n$). Taking into account \eqref{functiong}, the relation \eqref{eq:equicovma} for the quantum $\boldsymbol{\sigma}_{(n)}$ and classical $\boldsymbol{\sigma}_{(n)}^{\mathrm{cl}}$ covariance matrices, i.e., $\boldsymbol{\sigma}_{(n)}\simeq \boldsymbol{\sigma}_{(n)}^{\mathrm{cl}}$, and the Bohr-Sommerfeld quantization rule for the action variables, in the sense $ \hbar/2 \to I_{a}$, the classical analog of $g_\alpha$ is defined as
\begin{align}
    g^{\textrm{cl}}_\alpha \left(\tilde{\sigma}_k \right) :=  \frac{1}{ \left( \tilde{\sigma}_k+\frac{1}{2}\right)^\alpha -\left( \tilde{\sigma}_k-\frac{1}{2}\right)^\alpha}\,, \label{classicalfuncg}
\end{align}
where $\tilde{\sigma}_k$ is given by \eqref{eq:cleige}. Notice that \eqref{classicalfuncg} can be obtained from \eqref{functiong} just by replacing the $\nu_i$ with $\tilde{\sigma}_k$, i.e., we are only changing the domain of the function, just as we did in the definition of the function $\mathcal{S}^{\textrm{cl}}$ given by \eqref{indientro}.

With the help of the function $g^{\textrm{cl}}_\alpha$, we define {\it classical analogs} of the generalized purities, Bastiaans-Tsallis entropies, and R\'enyi entropies for $\alpha \geq 0$ as
\begin{subequations} \label{classicalanalogs}
\begin{align}
 \tilde{\mu}^{\textrm{cl}}_{\alpha} \left( a_1,a_2,\dots,a_n \right) &:= \prod_{k=1}^{n} g^{\textrm{cl}}_{\alpha} ( \tilde{\sigma}_k) \,, \label{eq:cagepu}\\
   \tilde{S}^{\textrm{cl}} _\alpha\left( a_1,a_2,\dots,a_n \right) &:=  \frac{1- \prod_{k=1}^{n} g^{\textrm{cl}}_{\alpha} ( \tilde{\sigma}_k) }{\alpha-1} \,, \label{eq:caenBT}\\
    \tilde{H}^{\textrm{cl}}_\alpha \left( a_1,a_2,\dots,a_n \right) &:=  \frac{\sum_{k=1}^{n} \ln \left( g^{\textrm{cl}}_{\alpha} ( \tilde{\sigma}_k) \right)}{1-\alpha} \,, \label{eq:caenR}
\end{align}
\end{subequations}
respectively.  We point out that the functions \eqref{classicalanalogs} do not need to invoke any prior knowledge from the quantum framework, and as we will see through examples, they yield exactly the same mathematical results as their quantum counterparts. Also, notice that the functions \eqref{classicalanalogs} are defined for any integrable system, but only when the corresponding quantum system is in a Gaussian state they correspond to the classical analogs of the generalized purities, Bastiaans-Tsallis entropies, and  R\'enyi entropies.

It is worth noticing that 
\begin{align}
    \lim_{\alpha \to 1} \frac{\sum_{k=1}^{n} \ln \left( g^{\textrm{cl}}_{\alpha} ( \tilde{\sigma}_k) \right)}{1-\alpha}&=  \lim_{\alpha \to 1} \frac{1- \prod_{k=1}^{n} g^{\textrm{cl}}_{p} ( \tilde{\sigma}_k) }{\alpha-1} \nonumber\\
    &= \sum_{k=1}^{n} \mathcal{S}^{\textrm{cl}}(\tilde{\sigma}_k) \,,
\end{align}
then
\begin{align}
    \lim_{\alpha \to 1}  \tilde{S}^{\textrm{cl}}_\alpha\left( a_1,a_2,\dots,a_n \right) &=  \lim_{\alpha \to 1}  \tilde{H}^{\textrm{cl}}_\alpha \left( a_1,a_2,\dots,a_n \right) \nonumber\\
    &=   \tilde{S}^{\textrm{cl}} \left( a_1,a_2,\dots,a_n \right) \,,
\end{align}
this is, in the limit $\alpha \to 1$ our classical analogs of the Bastiaans-Tsallis, $\tilde{S}^{\textrm{cl}}_\alpha$, and R\'enyi, $\tilde{H}^{\textrm{cl}}_\alpha$,  entropies become the classical analog of the von Neumann entropy \eqref{eq:caen}, in complete analogy with the quantum case.

\subsection{Logarithmic negativity for Gaussian states}

To establish a classical analog of the logarithmic negativity \eqref{LogNeg} (or \eqref{LogNeg2}), let us consider a classical integrable system of $N$  coupled harmonic oscillators.  As in the quantum case, we focus on a set of $m=n_{1}+n_{2}\leqslant N$ oscillators consisting of a group of $n_{1}$ oscillators and a group $n_{2}$ oscillators. Furthermore, we also assume that there are no correlations between positions and momenta. Under this considerations, we begin by introducing the reduced classical covariance matrix $\boldsymbol{\mu}^{\mathrm{cl}}$ associated with the $m$ oscillators, which subtracts the rows and columns corresponding to the $m$ oscillators from the  classical covariance matrix $\boldsymbol{\sigma}^{\mathrm{cl}}$. Then, the matrix $\boldsymbol{\mu}^{\mathrm{cl}}$ has the form 
\begin{equation}
  \boldsymbol{\mu}^{\mathrm{cl}}=\frac{1}{2}\left(\begin{array}{cc} \boldsymbol{\mu}^{\mathrm{cl}}_{q} & \boldsymbol{0}_{m \times m} \\ \boldsymbol{0}_{m \times m} & \boldsymbol{\mu}^{\mathrm{cl}}_{p} \end{array}\right), \label{redcovcl}
\end{equation}
where $\boldsymbol{\mu}^{\mathrm{cl}}_{q}$ and $\boldsymbol{\mu}^{\mathrm{cl}}_{p}$ are $m \times m$ matrices. Also in this case we can define the partial transpose $(\boldsymbol{\mu}^{\mathrm{cl}})^{\Gamma}$ of $\boldsymbol{\mu}^{\mathrm{cl}}$ with respect to the group of $n_{2}$ oscillators, namely 
\begin{equation}
 (\boldsymbol{\mu}^{\mathrm{cl}})^{\Gamma}:=\boldsymbol{P} \boldsymbol{\mu}^{\mathrm{cl}} \boldsymbol{P}\,, \label{covredcl}
\end{equation}
where $\boldsymbol{P}$ is given by \eqref{matrixP}. 

On the other hand, the classical analog of the symplectic matrix $\boldsymbol{\Omega}$, denoted by $\boldsymbol{\Omega}^{\mathrm{cl}}$, can be obtained by replacing the commutator $[\, , \, ]$ with the Poisson
brackets ${\rm i} \hbar \{\, , \, \}$. From this, we obtain the relation
\begin{equation}
  \boldsymbol{\Omega} \simeq\hbar \boldsymbol{\Omega}^{\mathrm{cl}}\,,  
\end{equation}
where $\boldsymbol{\Omega}^{\mathrm{cl}}=(\Omega^{\mathrm{cl}}_{\alpha \beta})$ is the classical symplectic matrix and has entries given by
\begin{equation}
\Omega^{\mathrm{cl}}_{\alpha \beta}:=\left\{r_{\alpha}, r_{\beta}\right\}\,.   
\end{equation}
Using $\{q_{a}, p_{b}\}=\delta_{ab}$, it is not hard to realize that 
\begin{equation}
\boldsymbol{\Omega}^{\mathrm{cl}}=\left(\begin{array}{ll}
\boldsymbol{0}_{N \times N} & \boldsymbol{1}_{N \times N} \\
-\boldsymbol{1}_{N \times N} & \boldsymbol{0}_{N \times N}
\end{array}\right).
\end{equation}
Hence, for the set of $m$ oscillators the reduced classical symplectic matrix $\boldsymbol{\Sigma}^{\mathrm{cl}}$ can be written as
\begin{equation}
\boldsymbol{\Sigma}^{\mathrm{cl}}=\left(\begin{array}{ll}
\boldsymbol{0}_{m \times m} & \boldsymbol{1}_{m \times m} \\
-\boldsymbol{1}_{m \times m} & \boldsymbol{0}_{m \times m}
\end{array}\right).\label{sympredcl}
\end{equation}

Using \eqref{covredcl} and \eqref{sympredcl}, it is natural to define a classical analog of the logarithmic negativity \eqref{LogNeg} as
\begin{equation}
   E^{\textrm{cl}}_{\mathcal{N}}=-\sum_{k=1}^{2 m} \log _{2}\left[\min \left(1, \mid \lambda^{\textrm{cl}}_{k} \mid \right)\right]\,, \label{LogNegCl}
\end{equation}
where $\lambda^{\textrm{cl}}_{k}$  are the eigenvalues of the matrix
\begin{equation}
    \boldsymbol{B}^{\textrm{cl}}=  \frac{\rm i}{2 c_4} \lim_{I_a \to c_4} (\boldsymbol{\Sigma}^{\textrm{cl}})^{-1} (\boldsymbol{\mu}^{\textrm{cl}})^{\Gamma}\,. \label{Bcl}
\end{equation}
Here, $c_4$ is an auxiliary real positive constant. Let us make some comments regarding \eqref{LogNegCl}. First, it is worth noticing that $E^{\textrm{cl}}_{\mathcal{N}}$ is a purely classical quantity since its definition does not require resorting to the quantum framework. Second, the (arbitrary) constant $c_4$ disappears from \eqref{Bcl} once the limit $I_a \to c_4$ is taken, and therefore $E^{\textrm{cl}}_{\mathcal{N}}$ does not depend on this constant. Third, by following a procedure completely analogous to the one leading to \eqref{LogNeg2} (see Ref.~\cite{Audenaert2002}), the classical analog of the logarithmic negativity can be written in terms of $\boldsymbol{\mu}^{\mathrm{cl}}_{q}$ and $\boldsymbol{\mu}^{\mathrm{cl}}_{p}$ as
\begin{equation}
  E^{\textrm{cl}}_{\mathcal{N}} =- \sum_{j=1}^{m} \log _{2}\left[\min \left(1, \tilde{\lambda}^{\textrm{cl}}_{j}\right)\right] \,, \label{LogNegCl2}  
\end{equation}
where $\tilde{\lambda}^{\textrm{cl}}_{j}$ are the eigenvalues of  the matrix
\begin{equation}
	 \boldsymbol{\tilde{B}}^{\textrm{cl}}= \frac{1}{(2 c_4 )^{2}} \lim_{I_a \to c_4} \boldsymbol{\mu}^{\textrm{cl}}_{q} \boldsymbol{P}_{P} \boldsymbol{\mu}^{\textrm{cl}}_{p} \boldsymbol{P}_{P}. \label{Ccl}
\end{equation}

Note that using $E^{\textrm{cl}}_{\mathcal{N}}$ we can also introduce a classical analog of the negativity $\mathcal{N}$. Bearing in mind Eq~\eqref{Negativity}, we define a classical analog of the negativity as
\begin{equation}
    \mathcal{N}^{\textrm{cl}}=\frac{2^{E^{\textrm{cl}}_{\mathcal{N}}}-1}{2}.
\end{equation}
Since $\mathcal{N}^{\textrm{cl}}$ is trivially related to $E^{\textrm{cl}}_{\mathcal{N}}$, in the next section we only present numerical checks of $E^{\textrm{cl}}_{\mathcal{N}}$.

\section{Examples}\label{examples}

In this section, we present some examples to illustrate the application of the proposed classical functions. In Subsection \ref{LCHO}, we compute the classical analogs of the generalized purities, Bastiaans-Tsallis entropies, and  R\'enyi entropies for a linearly coupled harmonic oscillator system, and compare them with their quantum counterparts for the ground state of the system.  In the Subsection \ref{GHOC2}, we consider the particular case of the generalized harmonic oscillator chain given by the Hamiltonian \eqref{Hcasoparticular}  and compare the generalized purities and entropies with their classical analogs. Finally, in Subsection \eqref{CLO}, for a circular lattice of oscillators in several configurations, we compute the classical analog of the logarithmic negativity \eqref{LogNegCl} and compare it with its quantum counterpart.

We will see in these examples that our classical approach provides exactly the same results as their quantum counterparts when the quantum counterpart of the system is in a Gaussian state.
 
\subsection{Linearly coupled harmonic oscillators}\label{LCHO}

Let us consider the system composed of two coupled harmonic oscillators described by the Hamiltonian
\begin{equation}
H(q,p)= \frac{1}{2} \left( p^2_1+p^2_2 + A q_1^2+B q_2^2+ C q_1 q_2 \right)\,, \label{H2ch}
\end{equation}
where $A,B$, and $C$ are real parameters such that $A,B >0$, $A \neq B$, and $4AB-C^2\geq 0$. This system has been widely used for the analysis  of quantum entanglement \cite{Kim_2005, Paz2008}, and one of its features is that it presents a very large quantum entanglement for certain  parameter values \cite{Jaeger2007, Makarov2018, Han99}. Recently, in \cite{DGGV2022} this system was also used to study the classical analogs of purity, linear quantum entropy, and von Neumann entropy, finding that these classical functions provide the same results as their quantum counterparts. Furthermore, this model was used to study the classical analog of the quantum geometric tensor \cite{Alvarez2019}. Interestingly and in contrast to the classical functions studied in \cite{DGGV2022}, it was shown that the classical metric tensor does not yield the full parameter structure of its quantum counterpart, the cause being a quantum ordering anomaly. 

Using the corresponding action-angle variables $(\varphi_a,I_a)$ ($a=1,2$), the components of the classical covariance matrix $\boldsymbol{\sigma}^{\textrm{cl}}$ of the system are  \cite{DGGV2022}
\begin{subequations}\allowdisplaybreaks \label{ejeccm}
\begin{align} 
& \sigma^{\mathrm{cl}}_{q q} =  \left(
\begin{array}{cc}
 \frac{I_1 \cos^2 \beta}{\omega_1}+\frac{I_2 \sin^2 \beta}{\omega_2} &  \left( \frac{I_2}{\omega_2}-\frac{I_1}{\omega_1} \right)\sin \beta \cos \beta \\
 \left( \frac{I_2}{\omega_2}-\frac{I_1}{\omega_1} \right)\sin \beta \cos \beta  &  \frac{I_1 \sin^2 \beta}{\omega_1}+\frac{I_2\cos^2 \beta}{\omega_2} \\
\end{array}
\right)\,,\\
& \sigma^{\mathrm{cl}}_{ p p}  =  \nonumber \\
& \left(
\begin{array}{cc}
 I_1 \omega_1 \cos^2 \beta+ I_2\omega_2 \sin^2 \beta        &  \left( I_2\omega_2-I_1\omega_1\right) \sin \beta \cos \beta \\
\left( I_2\omega_2-I_1\omega_1\right) \sin \beta \cos \beta &  I_1\omega_1 \sin^2 \beta+ I_2\omega_2 \cos^2 \beta \\
\end{array}
\right)\,, \\
&\sigma^{\mathrm{cl}}_{qp}= 
\mathbf{0}_{2 \times 2}\,,
\end{align}
\end{subequations}
where $\omega_1$ and $\omega_2$ are the normal frequencies
\begin{align}
    \omega_1:=\sqrt{A-\frac{C}{2} \tan \beta} \,, \qquad    \omega_2:=\sqrt{B+\frac{C}{2} \tan \beta} \,,
\end{align}
and $\tan 2 \beta = C/(B-A)$.

In this case, the subsystems are: each oscillator \{1\} and \{2\}; and the complete system \{1,2\}. For the subsystem $\{1\}$ the corresponding $\tilde{\sigma}_1$  [see \eqref{eq:cleige}] is
\begin{align}
\tilde{\sigma}_1 &=\frac{1}{2} \sqrt{\left(\frac{\cos^2 \beta }{\omega_1}+\frac{\sin ^2 \beta}{\omega_2}\right) \left(\omega_1 \cos^2 \beta+ \omega_2 \sin^2 \beta \right)} \label{Marcoseg}\,,
\end{align}
which in terms of the original parameters reduces to
\begin{align}
\tilde{\sigma}_1&= \sqrt{\frac{A B}{4A B- C^2}} \,, \label{eH2}
\end{align}
and, because of the symmetry between the subsystems $\{1\}$ and $\{2\}$, the symplectic eigenvalue of the subsystem $\{2\}$ satisfies $\tilde{\sigma}_2=\tilde{\sigma}_1$. Notice that in the uncoupled regime $C=0$, $\tilde{\sigma}_1=\tilde{\sigma}_2=1/2$ and then the classic analogs of the generalized purities and entropies are $1$ and $0$ for all $\alpha>1$, respectively, which means that the oscillators  are separated in phase space. Furthermore, the symplectic eigenvalues of the complete system $\{1,2\}$ are $ \sigma_1=1/2= \sigma_2=\tilde{\sigma}_1=\tilde{\sigma}_2$, which  means that the complete system is pure as expected. 

\begin{figure}[h!]
	\begin{tabular}{c c}
		\includegraphics[width=7.8cm]{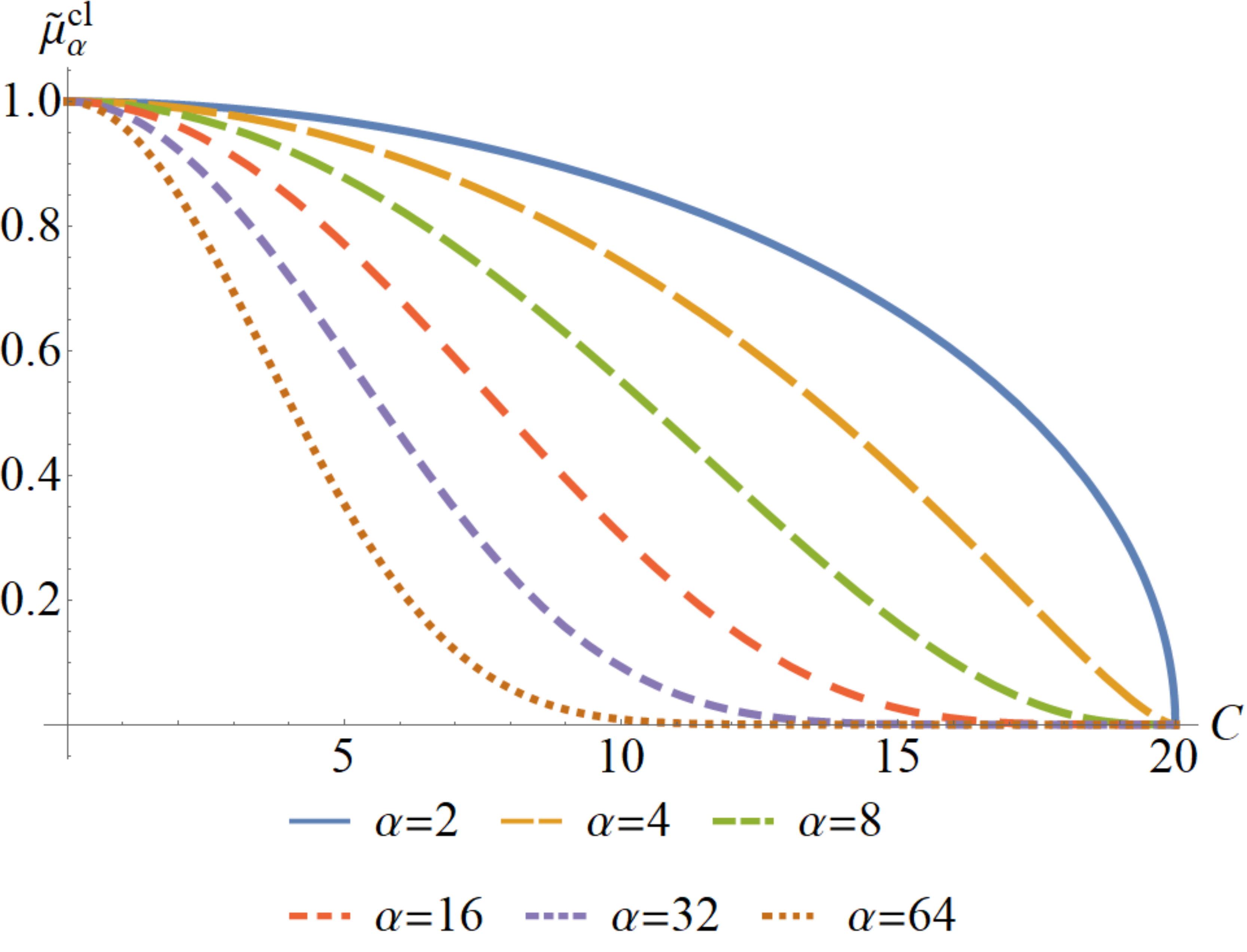} \\
		(a)\\
		
		  \includegraphics[width=7.8cm]{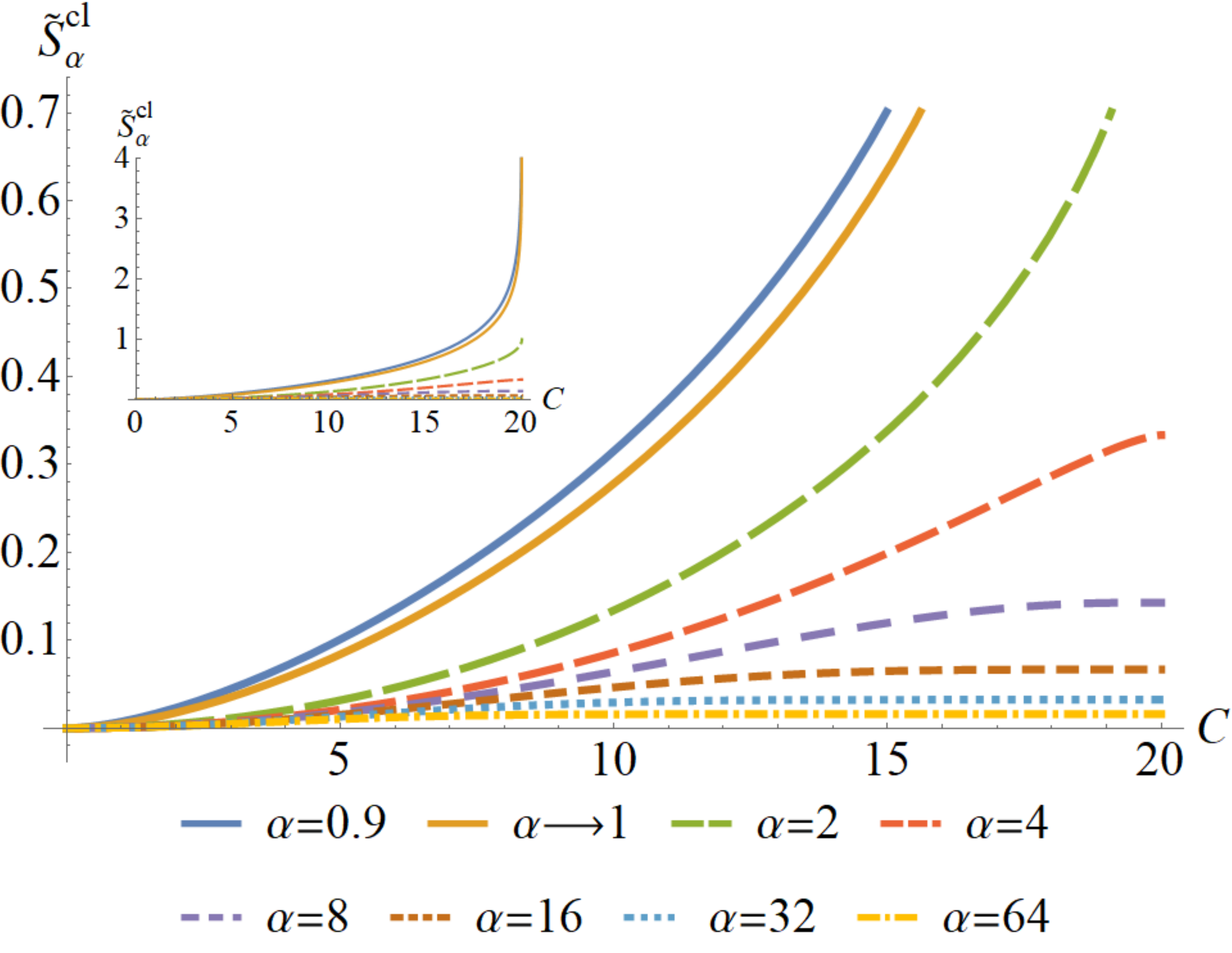}\\
		 (b) \\
  	\includegraphics[width=7.8cm]{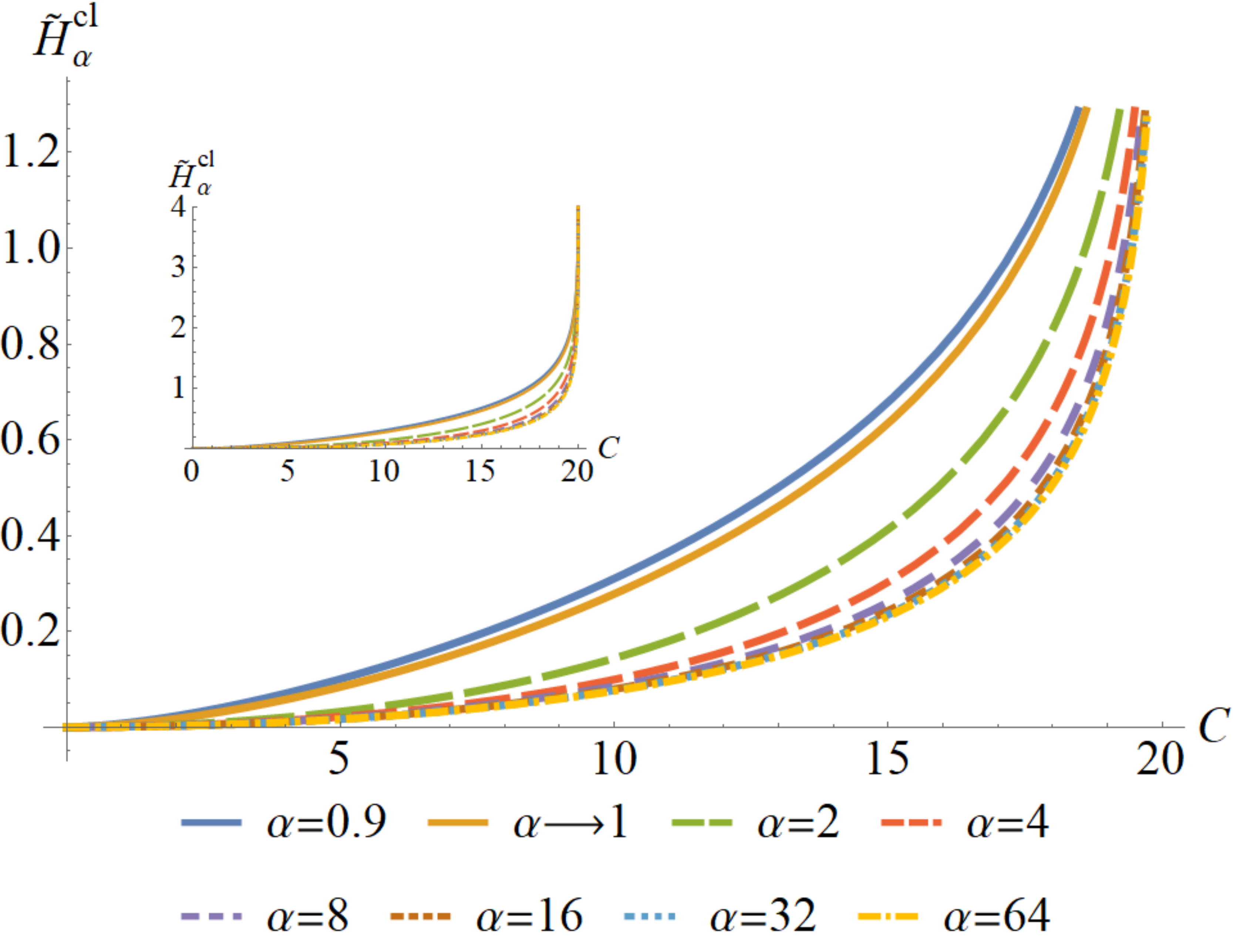} \\
		(c) 
	\end{tabular}
	\caption{Plots of the classical analogs, as a function of $C$ for different values $\alpha$,  of (a) generalized purities, (b) Bastiaans-Tsallis, and (c) R\'enyi entropies.
	} \label{purities1}
\end{figure}

Using \eqref{classicalanalogs} and  \eqref{eH2}, we compute the classical analogs of the generalized purities and entropies defined by \eqref{classicalanalogs}  for $\alpha \to 1$ and $\alpha = 0.9, 2,4,8,16,32,64$. Also, for illustrative purposes, we set $AB=100$. In Figs. \ref{purities1} we show the numerical results of the classical analogs as functions of the coupling constant $C$, which with the election $AB=100$ can take values in the interval $(-20,20)$. As these functions are symmetric under $C\to -C$, we only plot the region $[0,20)$.  Let us make some comments about it: i) These plots illustrate what we have said about the case $C=0$. ii) We have included the case $\alpha=2$, then the figure  \ref{purities1} (a) contains the analog of the purity. iii) We have included in Figs.  \ref{purities1} (b) and (c) the limit $\alpha \to 1$, which corresponds with the von Neumann entropy, given by 
\begin{subequations}\label{entropy-maka}
\begin{align}
\tilde{S}^{\mathrm{cl}}(1) =& \left(  \sqrt{\frac{A B}{4A B- C^2}}+\frac{1}{2} \right)\ln \left(  \sqrt{\frac{A B}{4A B- C^2}}+\frac{1}{2} \right)  \nonumber\\
&\!\!\!\!\!\!\!-\left(  \sqrt{\frac{A B}{4A B- C^2}}-\frac{1}{2} \right) \ln \left(  \sqrt{\frac{A B}{4A B- C^2}}-\frac{1}{2} \right)\nonumber\\
&=\tilde{S}^{\mathrm{cl}}(2)\,,\\
\tilde{S}^{\mathrm{cl}}(1,2)&=0 \,.
\end{align}
\end{subequations}
This classical analog was discussed in \cite{DGGV2022}. 
iv) In the Figs. \ref{purities1} (b) and (c)  we inset sub-figures to illustrate the global behavior of the classical analogs of the generalized entropies. v) We can appreciate some general characteristics of the generalized entropies; for instance, in Fig. \ref{purities1} (b)  we observe that Tsallis entropies are monotonically decreasing functions of $\alpha$ and  Fig. \ref{purities1} (c) shows the concave behavior of R\'enyi entropies.

Let us now consider the quantum case. Using the quantum counterpart of the Hamiltonian \eqref{H2ch}, the components of the quantum covariance matrix $\boldsymbol{\sigma}$ for the ground state  are
\begin{subequations}\allowdisplaybreaks \label{ejeqcm}
\begin{align}
 \sigma_{q q} &\!=  \! \frac{\hbar}{2} \!\left(
\begin{array}{cc}
 \frac{ \cos^2 \beta}{\omega_1}+\frac{ \sin^2 \beta}{\omega_2} &  \left( \frac{1}{\omega_2}-\frac{1}{\omega_1} \right)\sin \beta \cos \beta \\
 \left( \frac{1}{\omega_2}-\frac{1}{\omega_1} \right)\sin \beta \cos \beta  &  \frac{ \sin^2 \beta}{\omega_1}+\frac{\cos^2 \alpha}{\omega_2} \\
\end{array}
\right),\\
 \sigma_{ p p}  &\!= \! \frac{\hbar}{2}\! \left(
\begin{array}{cc}
  \omega_1 \cos^2 \alpha+ \omega_2 \sin^2 \beta        &  \left( \omega_2-\omega_1\right) \sin \beta \cos \beta \\
\left( \omega_2-\omega_1\right) \sin \beta \cos \beta &  \omega_1 \sin^2 \beta+ \omega_2 \cos^2 \beta \\
\end{array}
\right)\,,\\
\sigma_{qp}&\!= 
\mathbf{0}_{2 \times 2}\,.
\end{align}
\end{subequations}

Bearing in mind the Bohr-Sommerfeld quantization rule $I_a \to \hbar/2$, it is direct to show that the resulting classical and quantum covariant matrices, \eqref{ejeccm} and \eqref{ejeqcm}, are exactly the same. Therefore, the resulting  generalized purities and entropies \eqref{genequagaussian} of the subsystems are the same as determined from the classical functions \eqref{classicalanalogs}. This corroborates that our classical analogs are capable of giving rise to the same mathematical results as their quantum versions.

\subsection{Generalized harmonic oscillator chain}\label{GHOC2}

\begin{figure}[h!]
	\begin{tabular}{c c}
		\includegraphics[width=7.8cm]{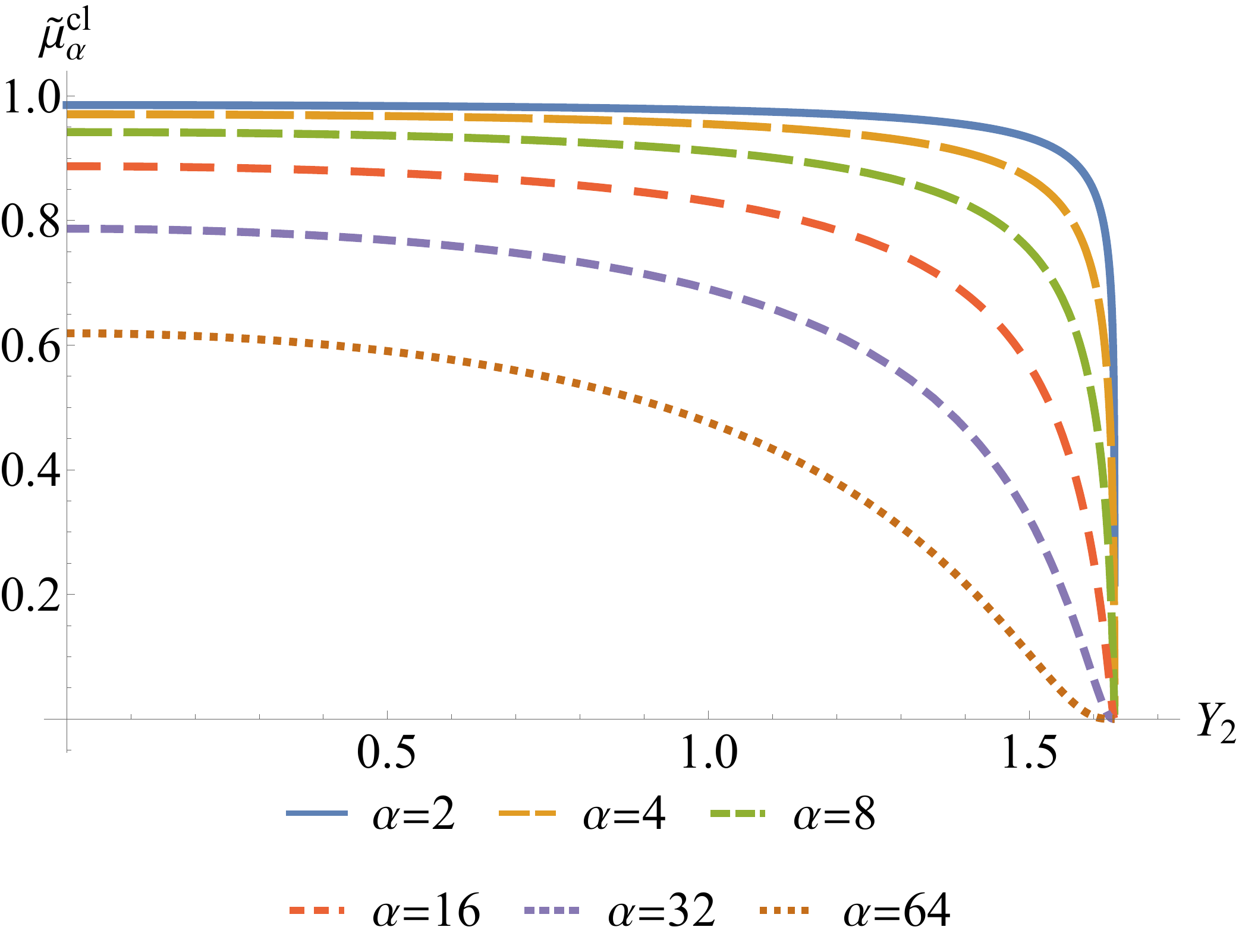} \\
		(a)\\
  
  \includegraphics[width=7.8cm]{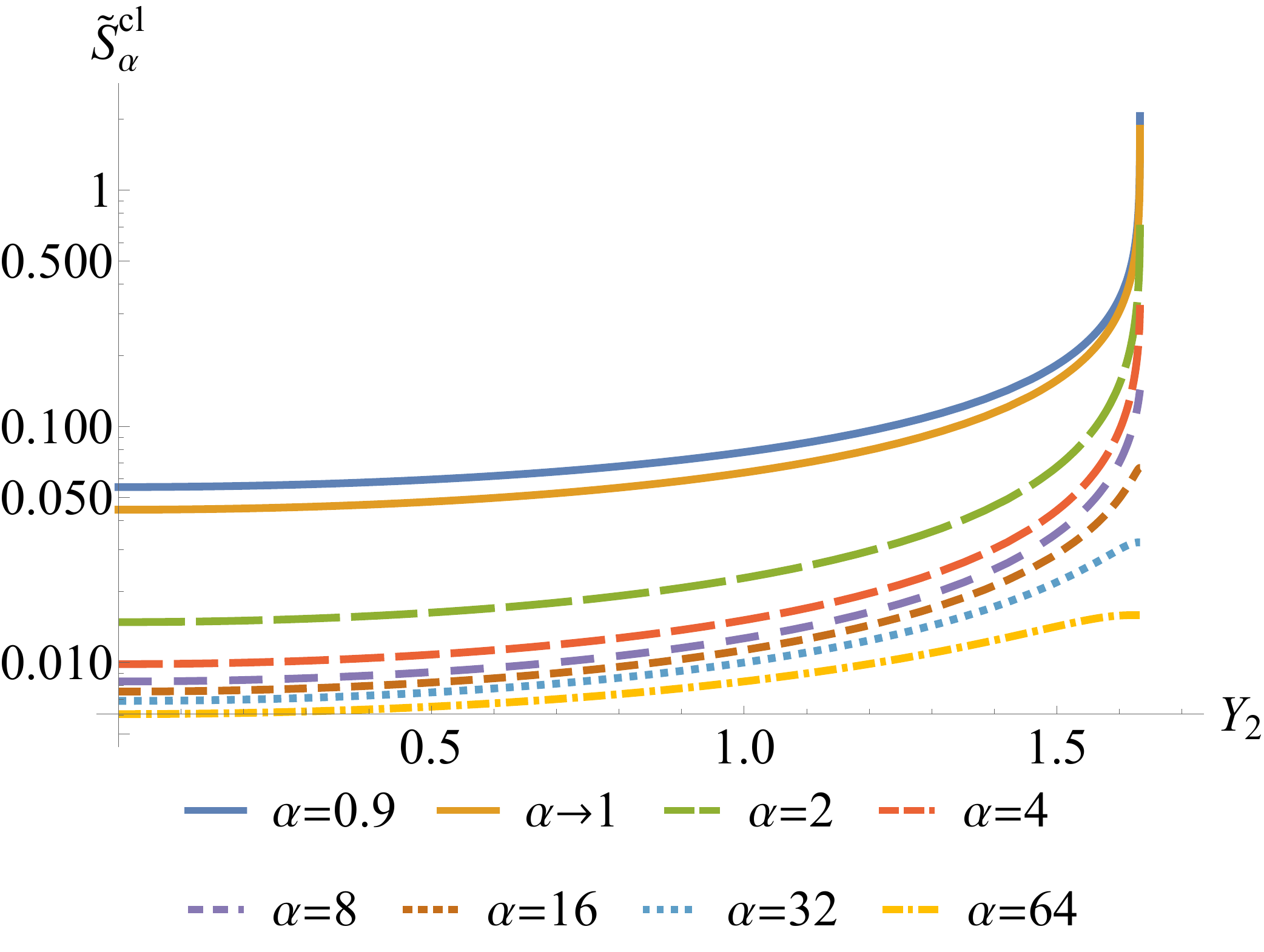} \\
		(b)\\	
		  \includegraphics[width=7.8cm]{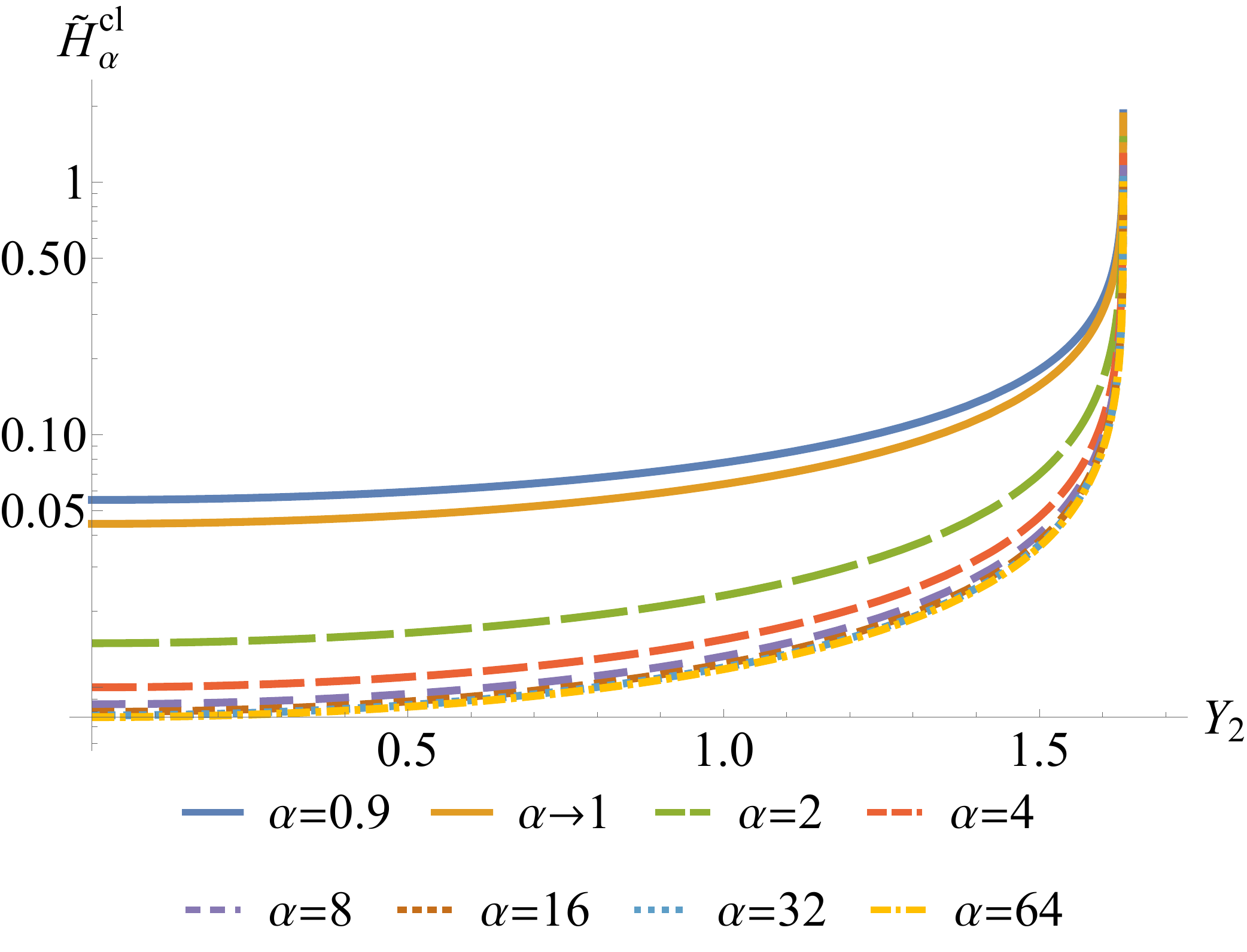}\\
		 (c)
	\end{tabular}
	\caption{Plots of the classical analogs, as a function of $Y_2$ for different values of $\alpha$, (a) generalized purities, (b)  Bastiaans-Tsallis and (c) R\'enyi entropies.
	} \label{puritiesgen}
\end{figure}

In this Subsection, we continue with the example of a generalized harmonic oscillator chain, whose Hamiltonian is presented in \eqref{Hcasoparticular}. For this system, we have already computed the classical covariance matrix and the classical analog of the purity. Now we will focus on the classical analogs of generalized purities, Bastiaans-Tsallis, and R\'enyi entropies, which can be calculated by using \eqref{classicalanalogs} and  the symplectic eigenvalues of the reduced covariance matrix \eqref{compccmex2}. The symplectic eigenvalues have the same functional form of \eqref{Marcoseg}, but with  normal frequencies given by \eqref{eq:frecuenciasgen}. The resulting expressions for these classical quantities will not be given explicitly, but instead, we plot them for $\alpha=0.9,2,4,8,16,32,64$ and $\alpha\to 1$, as functions of the parameter $Y_2$. We fix the rest of the parameters as $X_1=X_2=2$, $Z=1$, and $Y_1=0$.  This choice implies that $Y_2$ can take values in the interval $(-\sqrt{8/3},\sqrt{8/3})$ (out of this interval the normal frequencies have an imaginary part). As these functions are symmetric under $Y_2 \to -Y_2$ we only plot the region $[0, \sqrt{8/3})$. The results are shown in Fig. \ref{puritiesgen}. Notice that the classical analogs of R\'enyi entropies are always concave (see Fig. \ref{puritiesgen}.c), while the corresponding analogs of Bastiaans-Tsallis entropies are not (see Fig. \ref{puritiesgen}.b). On one hand, for $\alpha>1$, in the limit $Y_2 \to \sqrt{8/3}$ the classical analogs of generalized  purities go to zero (see Fig. \ref{puritiesgen}.a), while the classical analogs of R\'enyi entropies diverge and the  classical analogs of Bastiaans-Tsallis entropies go to $1/(\alpha-1)$. On the other hand, for $0 < \alpha \leq 1$ and $Y_2 \to \sqrt{8/3}$, the classical analogs of both R\'enyi and Bastiaans-Tsallis entropies diverge. Remarkably, we find that all these results (obtained by classical methods) coincide with those obtained by the quantum approach. 

\subsection{ One-dimensional circular lattice of oscillators}\label{CLO}

To close this section, we want to present numerical checks of our classical analog of the logarithmic negativity \eqref{LogNegCl2} for two groups of $n_{1}$ and $n_{2}$ oscillators in a system of $N$ identical oscillators on a one-dimensional circular lattice. The Hamiltonian of the system under consideration is
\begin{equation}
H=\frac{1}{2} \sum_{a=1}^{N} \left[p_a^2 + k q_a^2 + \kappa \left( q_a-q_{a+1} \right)^2 \right], \label{cirlat:H}
\end{equation}
with the periodic boundary condition $q_1\equiv q_{N+1}$. This system has been investigated in the context of logarithmic negativity \cite{Audenaert2002, Calabrese2012, Calabrese2013}, entanglement between collective operators \cite{Kofler2006}, and circuit complexity \cite{Jefferson2017}. It is worth noting that the Hamiltonian \eqref{cirlat:H} is a particular case of the Hamiltonian  \eqref{hamiltonianoqp}, i.e., the case where $\boldsymbol{Y}=0$ and $\boldsymbol{K}$ is a $N\times N$ symmetric circulant matrix given by
\begin{equation}
\renewcommand\arraystretch{1.7}
\boldsymbol{K}=\begin{pmatrix}
k+2\kappa & -\kappa & 0  &  \cdots  & -\kappa\\
-\kappa & k+2\kappa & -\kappa &   \cdots & 0\\
0 & -\kappa & \ddots &   \ddots &  \vdots\\
\vdots & \vdots & \ddots & \ddots & -\kappa\\
-\kappa & 0 & \cdots  &   -\kappa & k+2\kappa\\
\end{pmatrix} \, .
\end{equation}
This matrix is orthogonally diagonalizable and then can be expressed as $\boldsymbol{K}=\boldsymbol{U}^{\intercal}\boldsymbol{\Omega}^{2}\boldsymbol{U}$, where $\boldsymbol{U}$ is an orthogonal matrix and $\boldsymbol{\Omega}=\operatorname{diag}\left(\omega_{1}, \ldots, \omega_{N}\right)$ with $\omega_{1}, \ldots, \omega_{N}$ being the normal frequencies of the system. 

To study the classical analog of the logarithmic negativity, we now consider a set of $m=n_{1}+n_{2}\leqslant N$ oscillators consisting of two groups of $n_{1}$ and $n_{2}$ oscillators. In this case, the reduced $2m\times 2m$ classical covariance matrix $\boldsymbol{\mu}^{\mathrm{cl}}$ for the set of $m$ oscillators is obtained directly from the $2N\times 2N$ classical covariance matrix $\boldsymbol{\sigma}^{\textrm{cl}}$ of the system, 
\begin{equation}
	\boldsymbol{\sigma}^{\textrm{cl}}=\left(\begin{array}{cc} \boldsymbol{U}^{\intercal} \boldsymbol{\Omega}^{-1} \boldsymbol{I} \boldsymbol{U} & \boldsymbol{0}_{N \times N} \\ \boldsymbol{0}_{N \times N} & \boldsymbol{U^}{\intercal} \boldsymbol{\Omega} \boldsymbol{I} \boldsymbol{U} \end{array}\right) \, , \label{cirlat:covcl}
\end{equation}
by subtracting the corresponding rows and columns of the $m$ oscillators. Here, $\boldsymbol{I}=\operatorname{diag}\left(I_{1}, \ldots, I_{N}\right)$ is a diagonal matrix whose elements are the action variables $I_{a}$. From the resulting matrix $\boldsymbol{\mu}^{\mathrm{cl}}$ we can read off the matrices $\boldsymbol{\mu}^{\textrm{cl}}_{q}$ and $\boldsymbol{\mu}^{\textrm{cl}}_{p}$ and then construct the matrix $\boldsymbol{\tilde{B}}^{\textrm{cl}}$ given by \eqref{Ccl} for each considered case.  It is important to point out that the resulting matrix $\boldsymbol{\tilde{B}}^{\textrm{cl}}$ does not depend on the  auxiliary constant $c_4$, despite the fact that its construction involves such a constant. Having $\boldsymbol{\tilde{B}}^{\textrm{cl}}$, we find the classical logarithmic negativity \eqref{LogNegCl2} by a simple numerical calculation.

\begin{figure}
	\includegraphics[width=7cm]{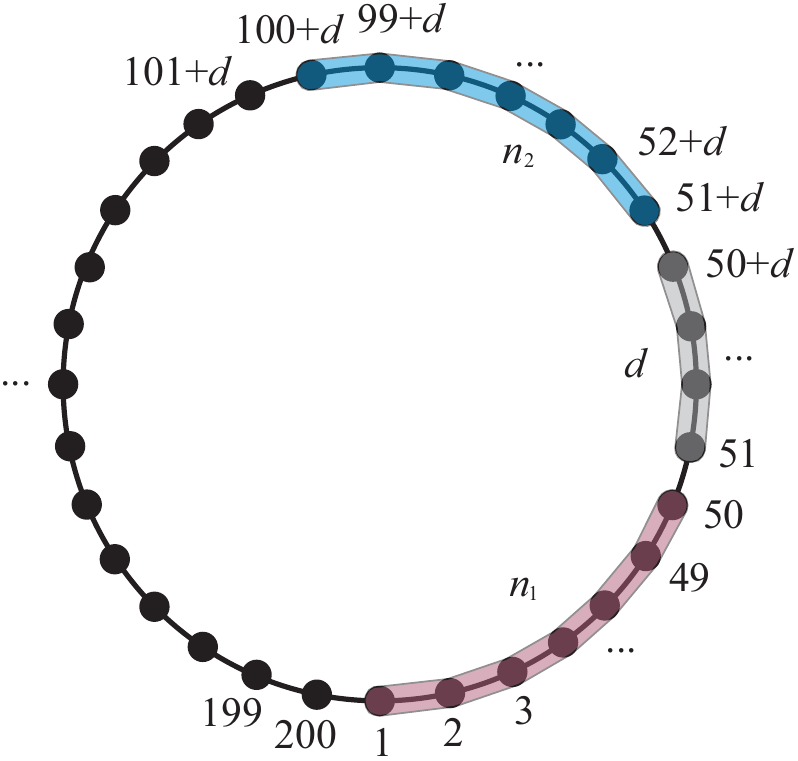}
	\caption{Illustration of the setup used in the first case. Two disjoint groups of $n_{1}=50$ and $n_{2}=50$ oscillators of a circular lattice of $N=200$ oscillators. The groups are separated by $d$ oscillators.} \label{Fig:LNconf1}
\end{figure}

Let us begin by considering a system of $N=200$ oscillators which involves two disjoint groups of $n_{1}=50$ and $n_{2}=50$ oscillators, separated by $d$ oscillators. The oscillators $51$ to $50+d$ and $101+d$ to 200 are not part of the two groups. In Fig. \ref{Fig:LNconf1}, we show an illustration of this setup.

\begin{figure}[H]
	\begin{tabular}{c c}
		\includegraphics[width=8cm]{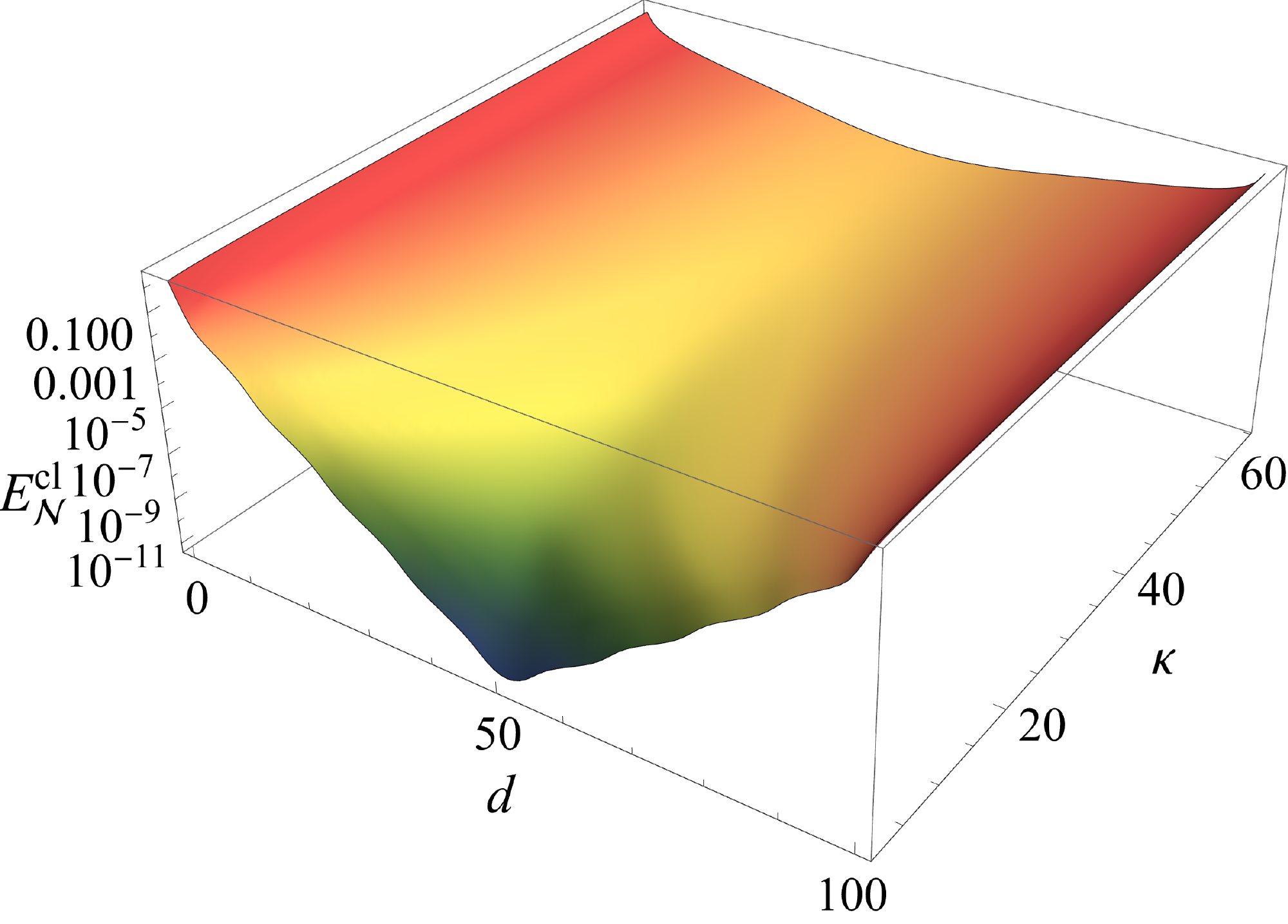} \\ 
		(a)\\
		 \textbf{\includegraphics[width=8cm]{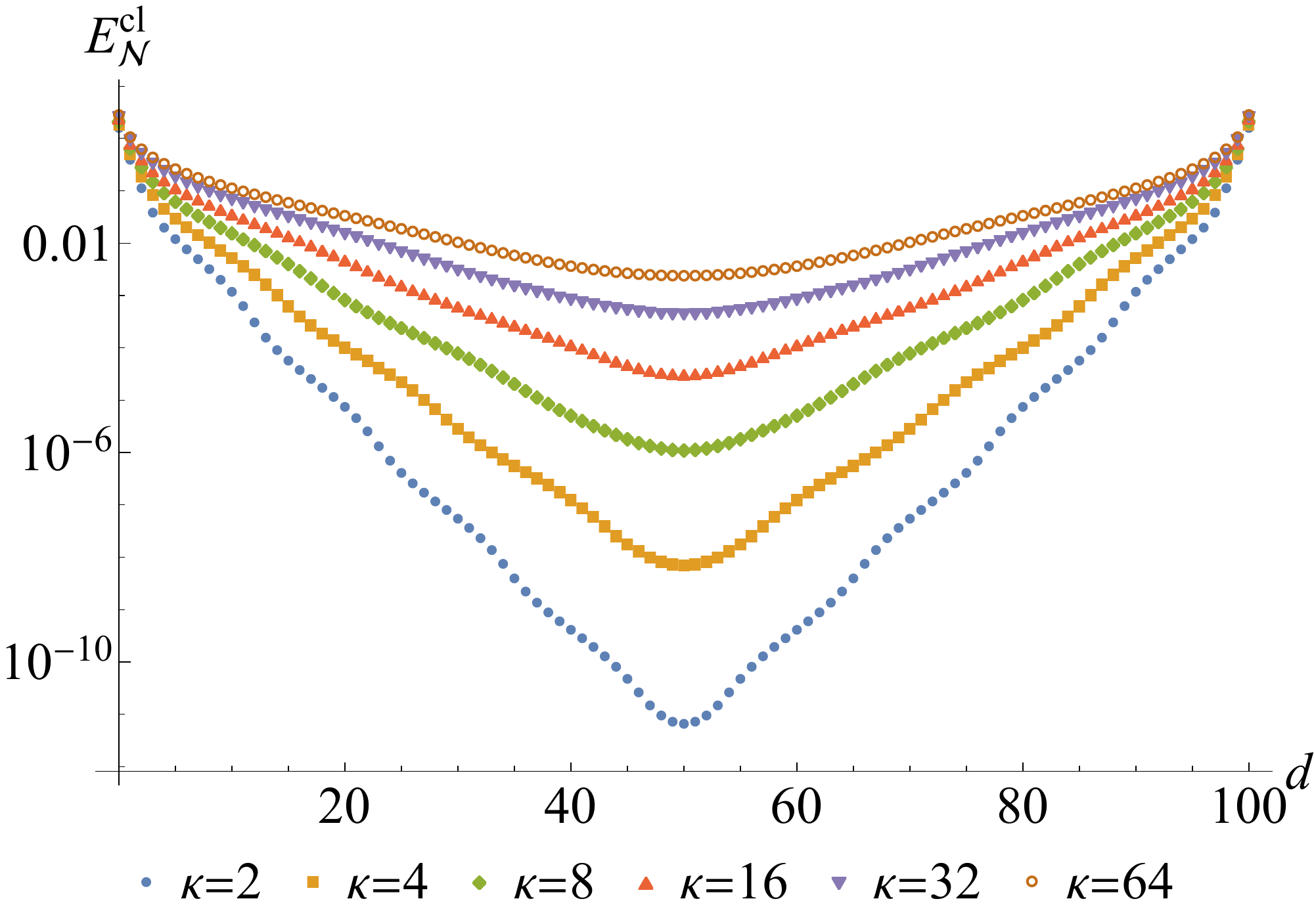}}\\
		 (b) \\
		\includegraphics[width=8cm]{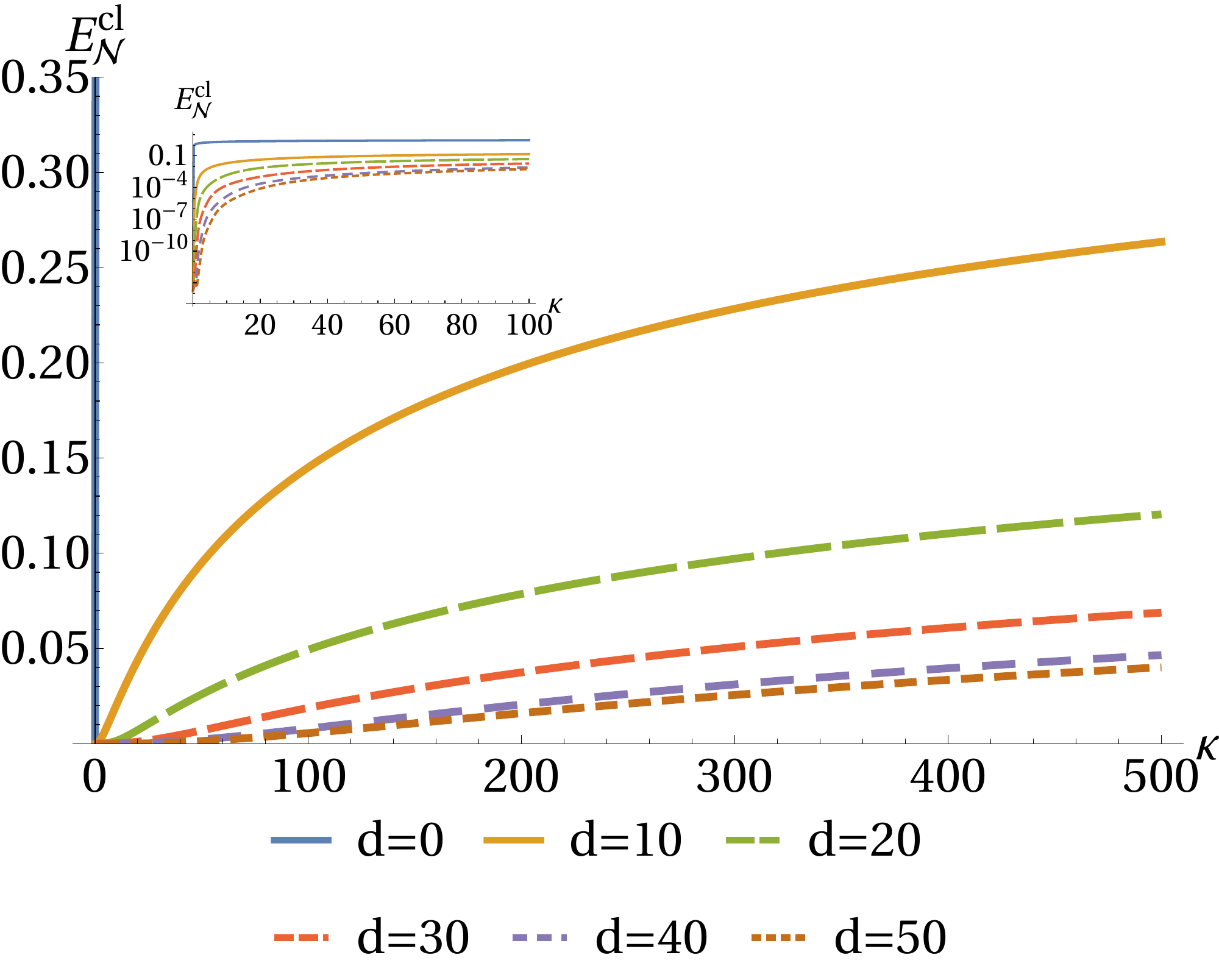} \\
		(c)  
	\end{tabular}
	\caption{Classical analog of the logarithmic negativity for two groups of $n_1=n_2=50$ oscillators embedded in a harmonic chain of $N=200$ oscillators with $k=0.1$. $d$ is the number of oscillators between the two groups, as shown in Fig. \ref{Fig:LNconf1}.} \label{Fig:LN1}
\end{figure}

In Fig.~\ref{Fig:LN1}(a), we plot the classical analog of the logarithmic negativity, on a logarithmic scale, as a function of $d$ and the coupling constant $\kappa$, with $k=0.1$. From this figure we see that, given a value of $\kappa$, the classical analog of the logarithmic takes a maximum value at $d=0$ and $d=100$, whereas it takes a minimum value when $d=50$ which corresponds to the most symmetric case. This can be better appreciated in Fig.~\ref{Fig:LN1}(b) where we plot the classical analog of the logarithmic negativity, on a logarithmic scale, as a function of $d$ for some values of $\kappa$. In this plot, we can also see that $E^{\textrm{cl}}_{\mathcal{N}}$ decreases from $d=0$ to $d=50$ following a near-exponential
form. This result is in agreement with the one reported in \cite{Audenaert2002} for the effect of the group separation on the logarithmic negativity. Notice that due to the symmetry of the setup,  $E^{\textrm{cl}}_{\mathcal{N}}$ increases from $d=50$ to $d=100$ following a near-exponential form. Another feature that we can see is that the minimum values of $E^{\textrm{cl}}_{\mathcal{N}}$ (at $d=50$) increase as $\kappa$ increases. From Fig.~\ref{Fig:LN1}(c) we can see the behavior of the classical analog of the logarithmic negativity with respect to $\kappa$, for different values of $d$. Here we see that $E^{\textrm{cl}}_{\mathcal{N}}$ increases as $\kappa$ increases, which is an expected result. It can be verified that these results coming from the classical analog of the logarithmic negativity are exactly the same as those obtained using logarithmic negativity given by Eq.~\eqref{LogNeg2}.

\begin{figure}[H]
	\includegraphics[width=7cm]{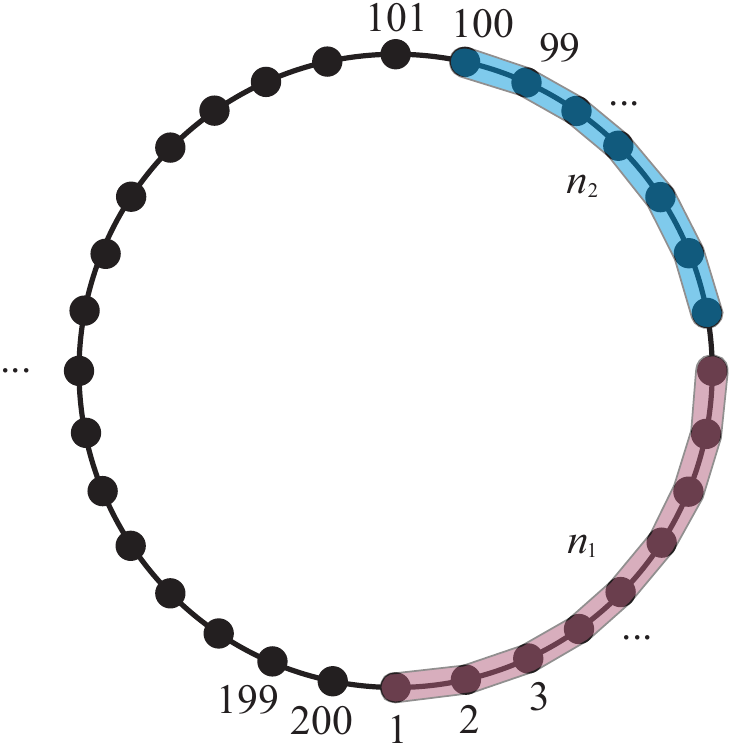}
	\caption{Illustration of the setup considered in the second case. Two adjacent groups of $n_{1}$ and $n_{2}=100-n_1$ oscillators of a circular lattice of $N=200$ oscillators.}\label{Fig:LNconf2}
\end{figure}

Let us now illustrate our approach on a system with two adjacent ($d=0$) groups of $n_{1}\leq100$ and $n_{2}=100-n_1$ oscillators in a circular lattice of $N=200$ oscillators. Note that in general the groups have  different sizes ($n_{1}\neq n_{2}$) while the set has a fixed size ($m=n_{1}+n_{2}=100$) and that the oscillators $101$ to $200$ are not part of the two groups. The setup is depicted in Fig. \ref{Fig:LNconf2}. In this case, we fix $k=0.0001$.

\begin{figure}[H]
	\begin{tabular}{c}
		\includegraphics[width=8cm]{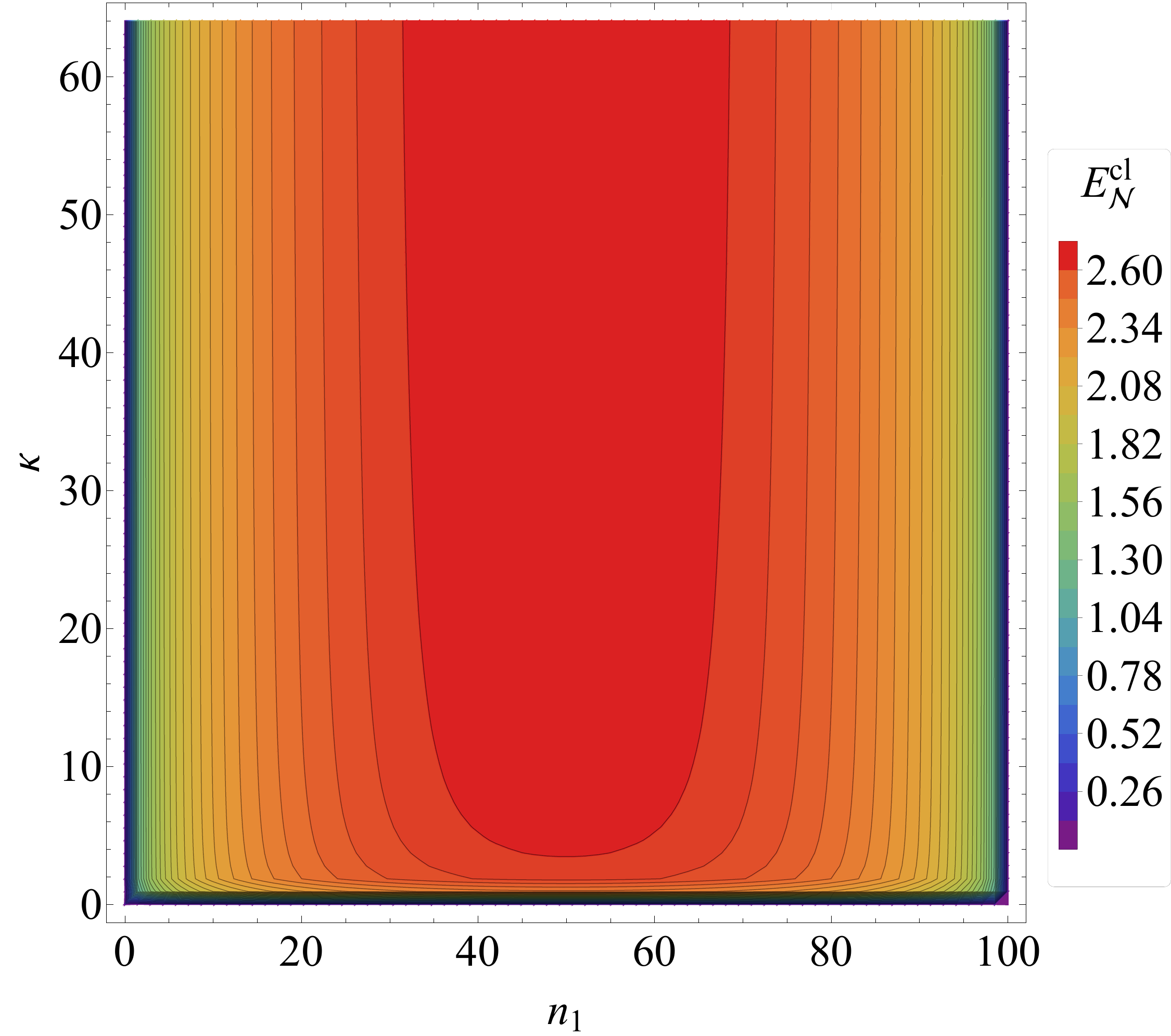} \\
		(a)\\
		\textbf{\includegraphics[width=8cm]{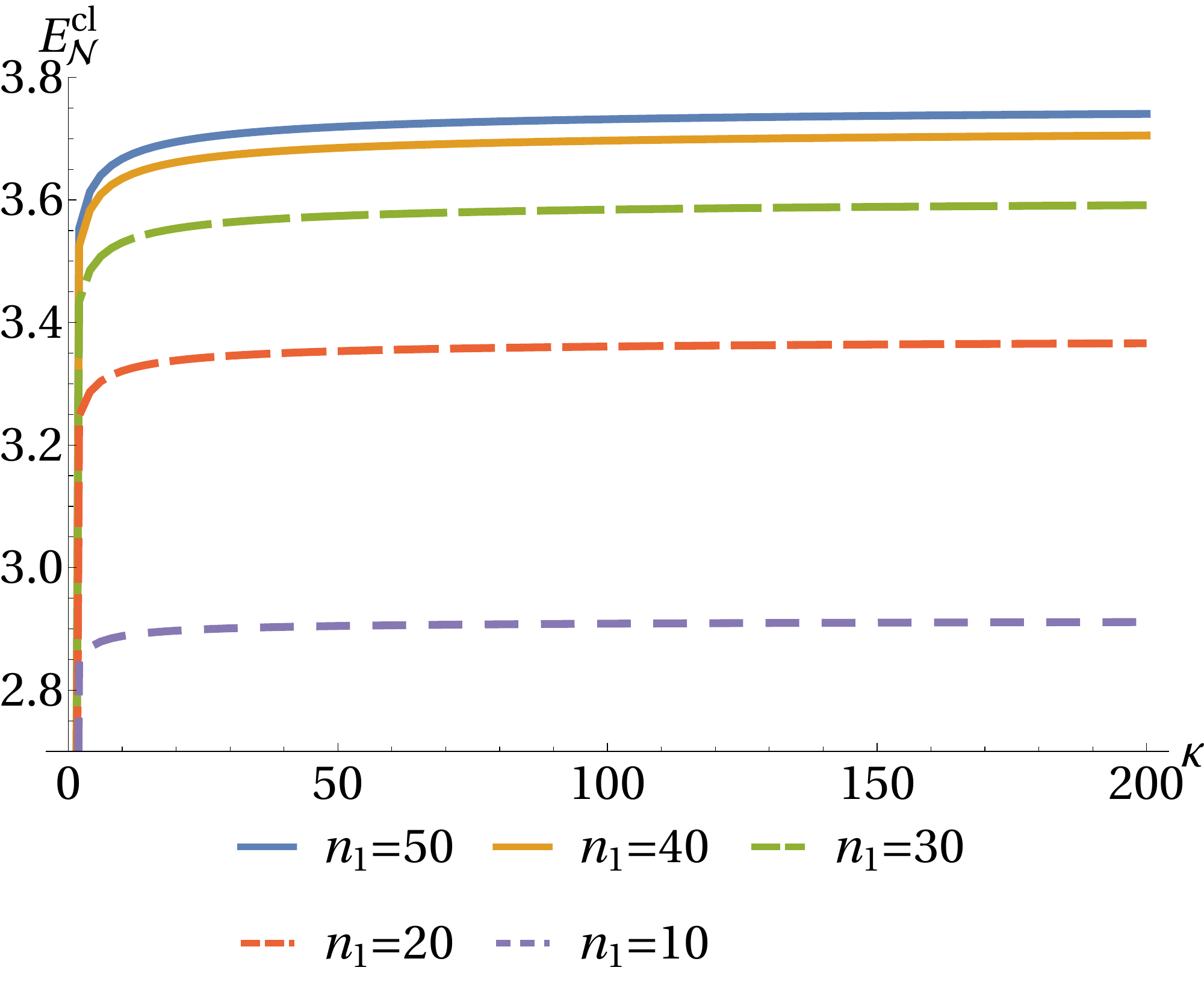}}\\
	 (b) \\
	 \textbf{\includegraphics[width=8cm]{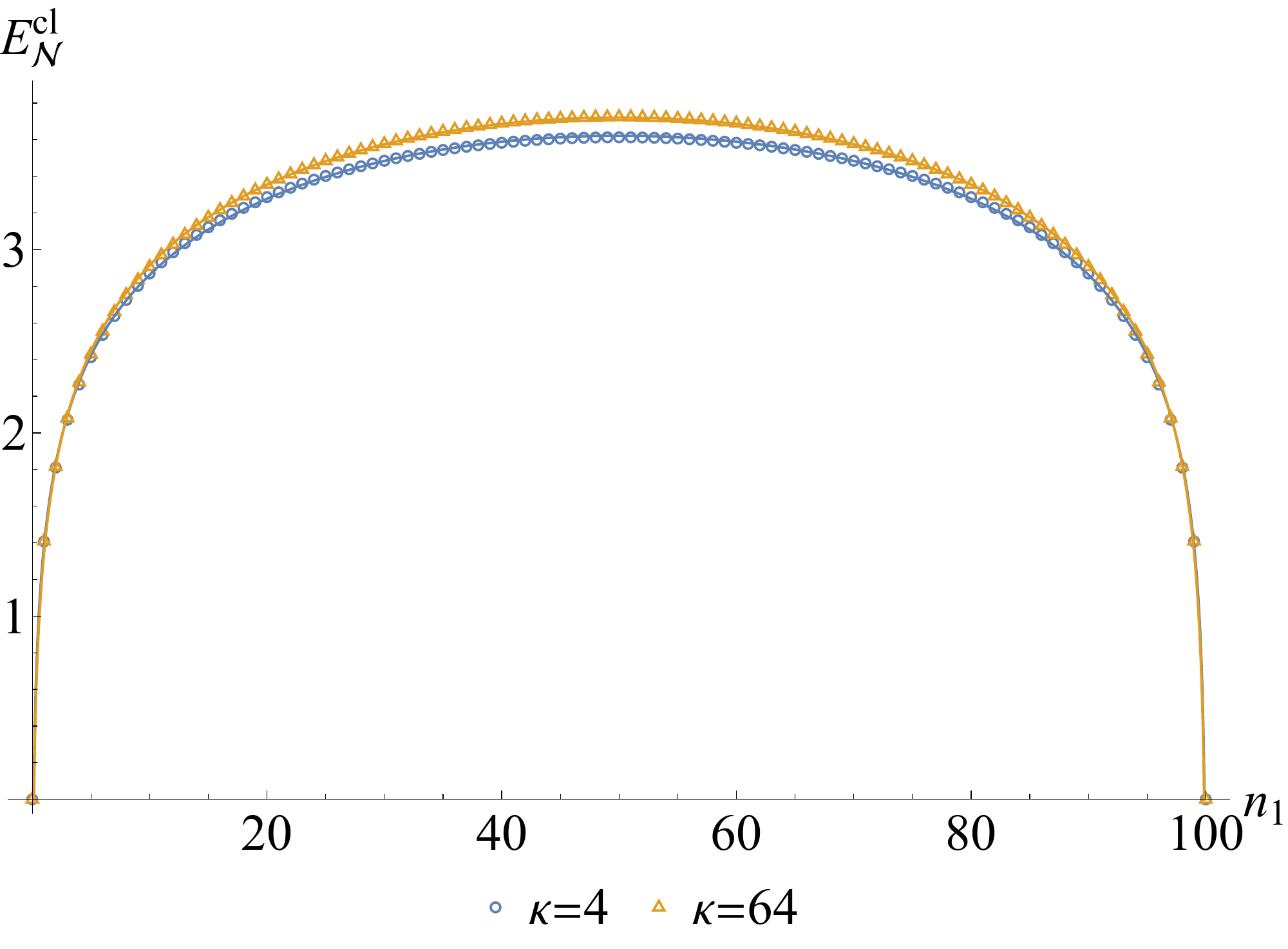}}\\
	 (c)
	\end{tabular}
	\caption{Classical analog of the logarithmic negativity for two adjacent groups of sizes $n_1$ and $n_2=100-n_1$ in a harmonic chain of $N=200$ oscillators with $k=0.0001$. Continuous lines in (c) are the fit obtained with \eqref{LNfinite}.}\label{Fig:LN2}
\end{figure}

The numerical calculation yields the results depicted in Fig. \ref{Fig:LN2}(a), where we show a map of the classical analog of the logarithmic negativity as a function of $\kappa$ and $n_1$. Note that when $n_1$ is kept fixed the classical function $E^{\textrm{cl}}_{\mathcal{N}}$ increases with $\kappa$. This behavior can be seen more clearly in Fig. \ref{Fig:LN2}(b), which also shows that the value of $E^{\textrm{cl}}_{\mathcal{N}}$ for $n_1=50$ $(n_1=n_2)$ is greater than the corresponding values for $n_1\neq50$ $(n_1\neq n_2)$. In Fig \ref{Fig:LN2}(c), we plot  the classical analog of the logarithmic negativity as a function of $n_1$ for $\kappa=2,64$. We see that the lowest value of $E^{\textrm{cl}}_{\mathcal{N}}$ is obtained for $n_1=0$ ($n_2=100$) and $n_1=100$ ($n_2=0$), while the highest value of $E^{\textrm{cl}}_{\mathcal{N}}$ is obtained for the symmetric configuration $n_1=50(=n_2)$. This effect of the group sizes $n_1$ and $n_2$ on $E^{\textrm{cl}}_{\mathcal{N}}$ is analogous to the one found in \cite{Audenaert2002} for asymmetrically bisected chains, in the sense that the value of $E^{\textrm{cl}}_{\mathcal{N}}$ with $n_1=n_2$ provides an upper bound on the values with $n_1\neq n_2$. On the other hand, the logarithmic negativity for two adjacent intervals of lengths $n_1$, $n_2$ of a finite system of length $N$  is given by \cite{Calabrese2012,Calabrese2013}
\begin{equation}\label{LNfinite2}
  E_{\mathcal{N}}=\frac{b_1}{4} \ln\left[\frac{N}{\pi} \frac{\sin\left(\frac{\pi n_1}{N} \right)\sin\left(\frac{\pi n_2}{N} \right)}{\sin\left(\frac{\pi (n_1+n_2)}{N} \right)}\right]+b_2,
\end{equation}
where $b_1$ and $b_2$ are constants. Plugging $N=200$ and $n_{2}=100-n_1$ into this expression, the logarithmic negativity reduces to
\begin{equation}\label{LNfinite}
  E_{\mathcal{N}}=\frac{b_1}{4} \ln\left[\frac{100}{\pi} \sin\left(\frac{\pi n_1}{100} \right) \right]+b_2.
\end{equation}

In Fig \ref{Fig:LN2}(c), we also plot $E_{\mathcal{N}}$ given by \eqref{LNfinite} as a function of $n_1$ with $b_1=2.5834$, $b_2=1.3864$ (blue continuous line) and $b_1=2.7464$, $b_2=1.3448$ (orange continuous line), which fit well to the data for $\kappa=4$ and $\kappa=64$, respectively. This corroborates the predictions made by our classical approach. Furthermore, we verify that the numerical results obtained in this case using the classical function $E^{\textrm{cl}}_{\mathcal{N}}$ are the same as those determined with the logarithmic negativity $E_{\mathcal{N}}$. It is interesting to note that, in the quantum context, $b_1$ in \eqref{LNfinite} (or \eqref{LNfinite2}) is the central charge which can be regarded as a measure of the degrees of freedom of a system. This means that the values of $b_1$ obtained by following our classical approach correspond to the central charge, since in this case, $E^{\textrm{cl}}_{\mathcal{N}}$ leads to the same results as its quantum counterpart. In general,  $E^{\textrm{cl}}_{\mathcal{N}}$ provides the same results as $E_{\mathcal{N}}$ for Gaussian states, and then the classical approach can also be used to compute the central charge. Nevertheless, understanding the role of the constant $b_1$ in the classical framework requires further analysis.

\begin{figure}[H]
	\includegraphics[width=7cm]{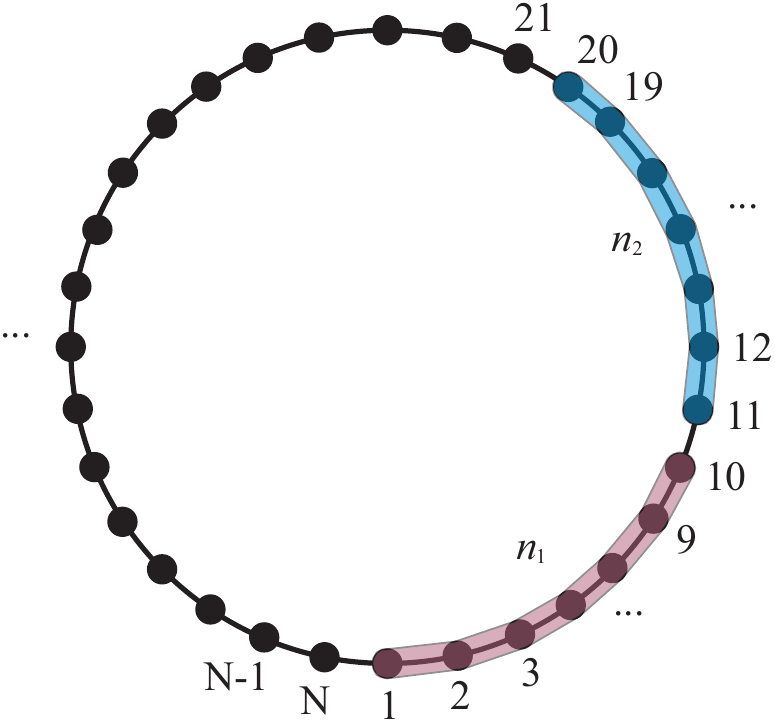}
	\caption{Illustration of the setup used in the third case. Two adjacent groups of $n_{1}=10$ and $n_{2}=10$ oscillators of a circular lattice of $N$ oscillators.}\label{Fig:LNconf3}
\end{figure}

Finally, the third case is devoted to studying two adjacent groups of equal sizes $n_1=n_2=10$ embedded in a harmonic chain of $N$ oscillators. The setup is illustrated in Fig. \ref{Fig:LNconf3}, where it is also clear that the oscillators $21$ to $N$ do not belong to the two groups.

\begin{figure}[H]
	\begin{tabular}{c}
		\includegraphics[width=8cm]{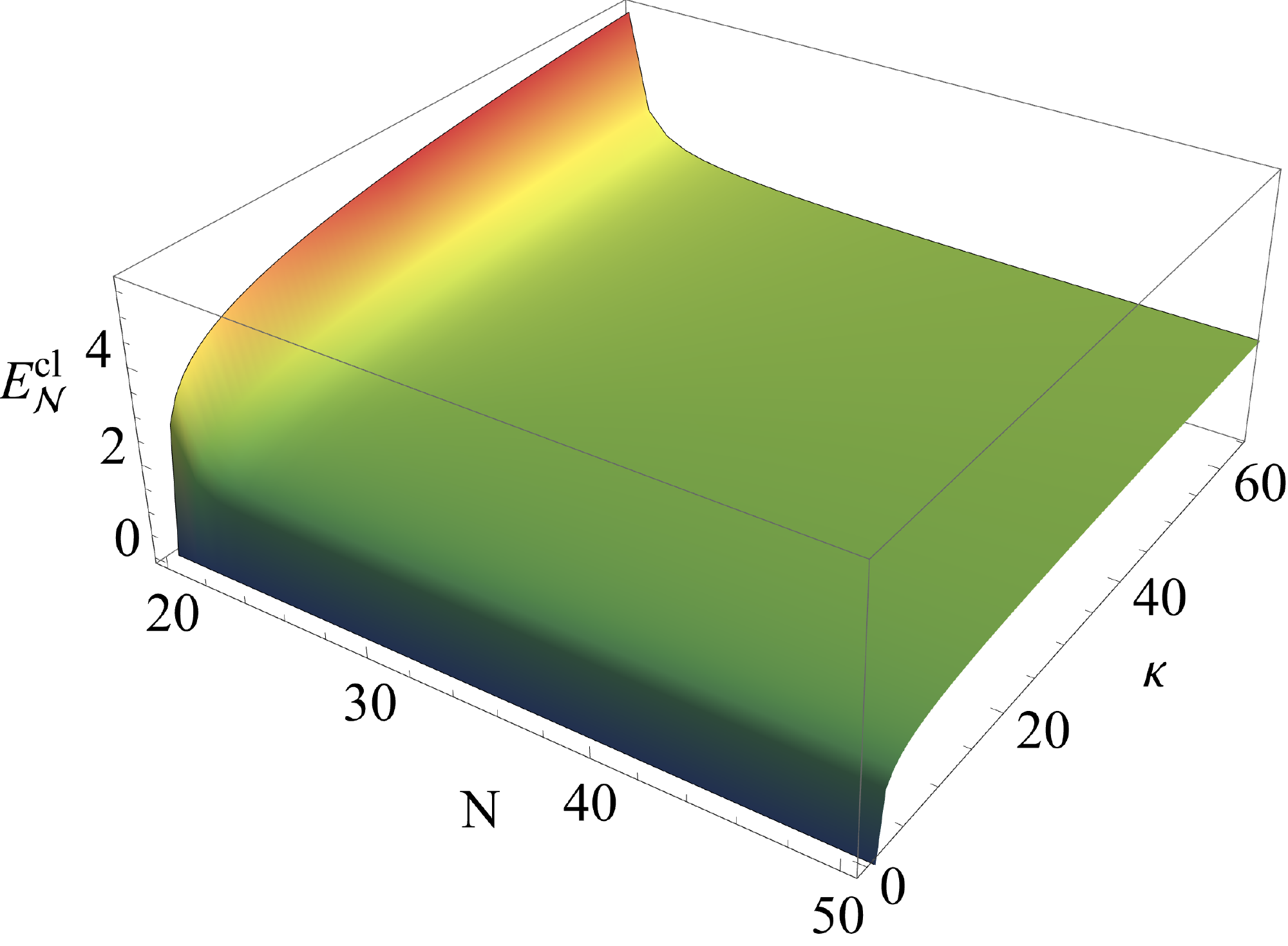} \\
		(a) \\
		 \textbf{\includegraphics[width=8cm]{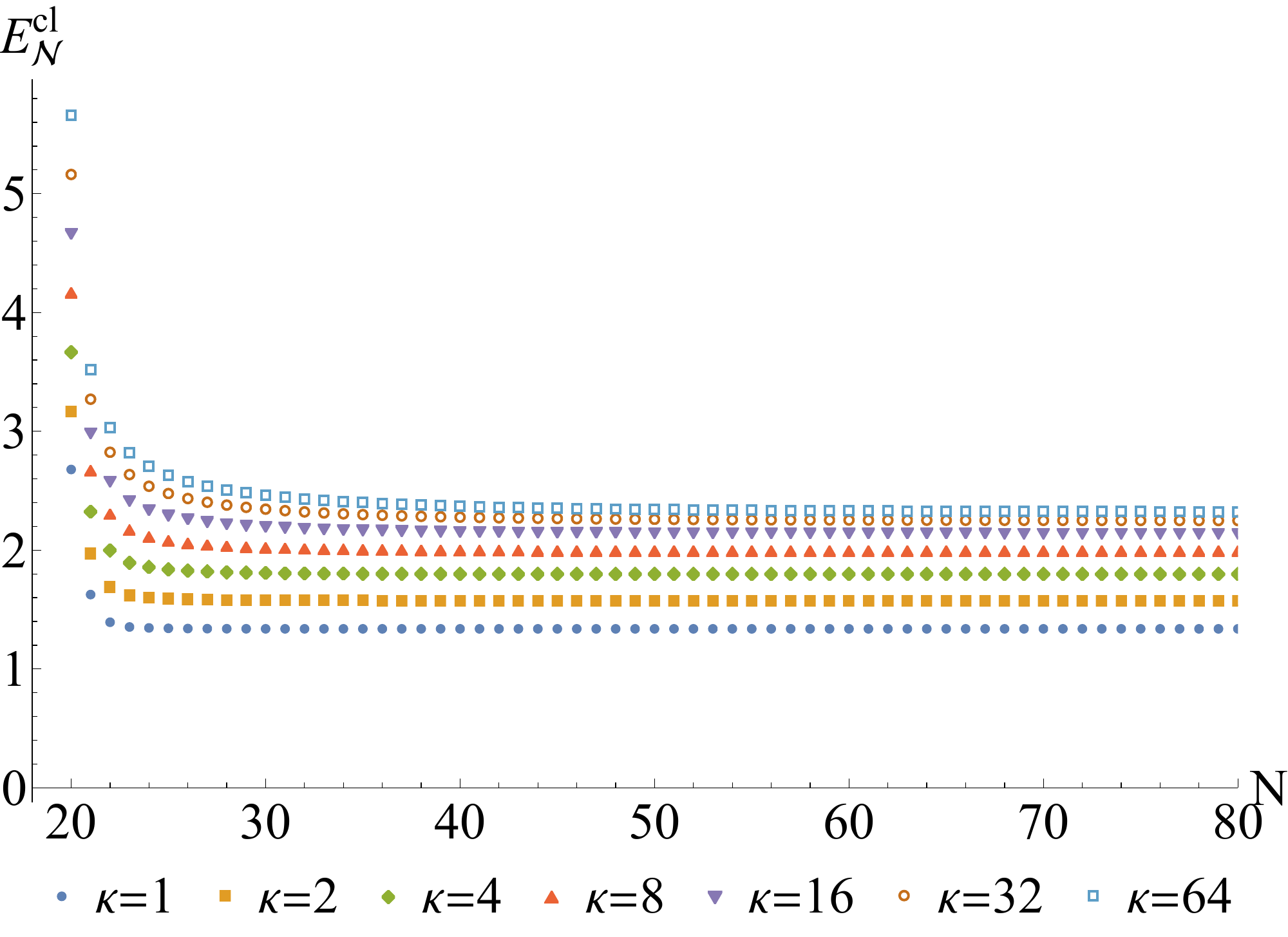}}\\
	    (b) 
	\end{tabular}
\caption{Classical analog of the logarithmic negativity for two adjacent groups of identical size $n_1=n_2=10$ in a harmonic chain of $N$ oscillators with $k=0.1$.} \label{Fig:LN3}
\end{figure}

In Fig. \ref{Fig:LN3}(a), we show the numerical results for the classical analog of the logarithmic negativity as a function of $N$ and $\kappa$. From this plot, we see two features. First, $E^{\textrm{cl}}_{\mathcal{N}}$ increases with $\kappa$, which also happened in the two previous cases and is expected. Second, for a fixed value of $\kappa$, $E^{\textrm{cl}}_{\mathcal{N}}$ decreases as $N$ increases and asymptotically approaches a finite constant value for large $N$. This behavior is shown explicitly in Fig. \ref{Fig:LN3}(b) for different values of the coupling constant $\kappa$. Moreover, from this plot, we can see that there is a dependence on $\kappa$ of the value to which $E^{\textrm{cl}}_{\mathcal{N}}$ approaches.

\begin{figure}[H]
	\begin{tabular}{c}
		\includegraphics[width=8cm]{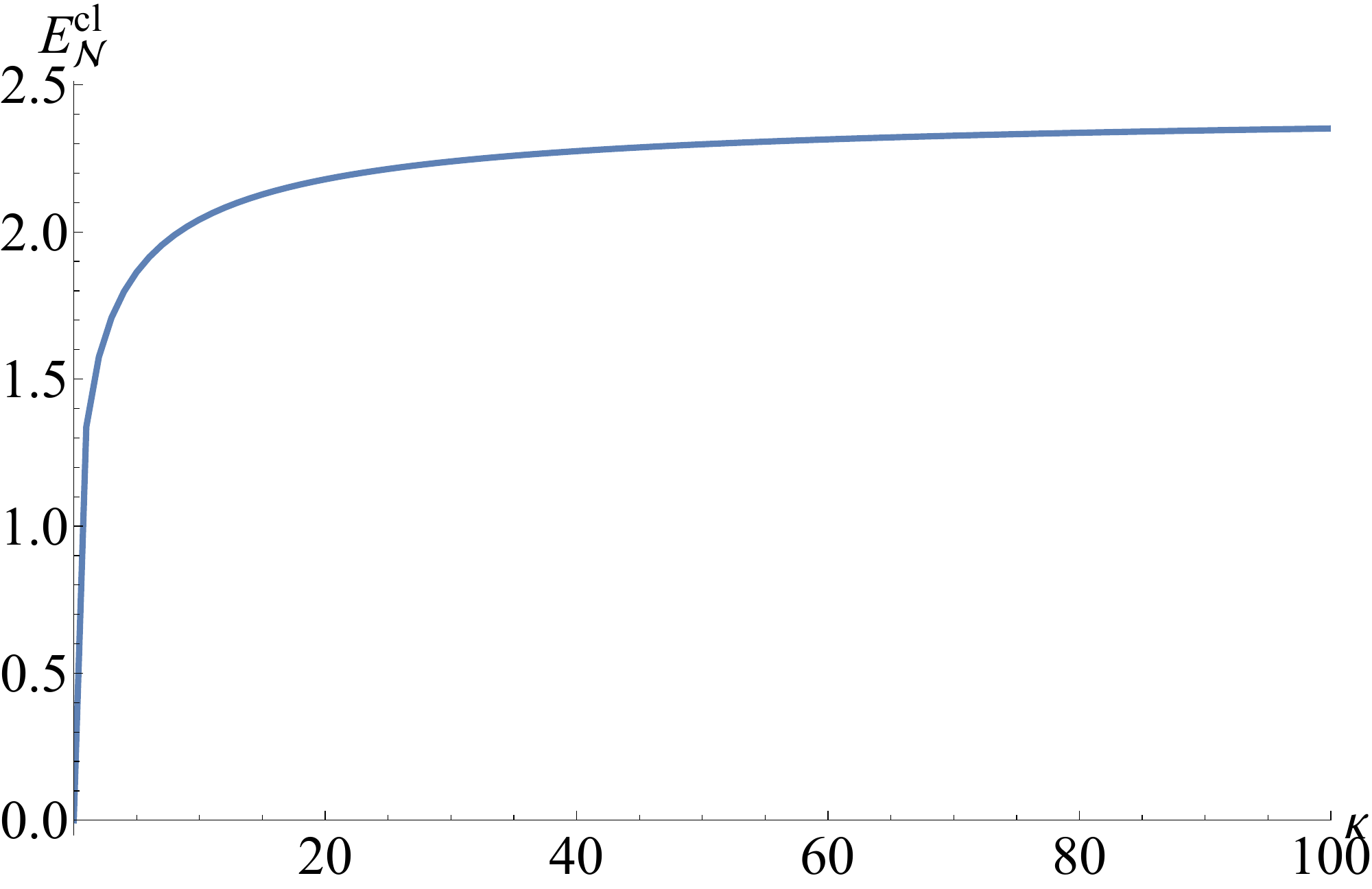}
	\end{tabular}
\caption{Classical analog of the logarithmic negativity for two adjacent groups of size $n_1=n_2=10$ in a fixed chain of $N=500$ oscillators.} \label{Fig:LN4}
\end{figure}

To gain more insight into this, in Fig. \ref{Fig:LN4} we show $E^{\textrm{cl}}_{\mathcal{N}}$ as a function of $\kappa$ for a fixed system size of $N=500$ oscillators, which in this case is large enough to avoid finite-size effects (see Fig. \ref{Fig:LN3}(b)). We see that, for large $N$, $E^{\textrm{cl}}_{\mathcal{N}}$ increases as $\kappa$ increases and behaves as
\begin{equation}
    E^{\textrm{cl}}_{\mathcal{N}} \sim 2.458-\frac{2.149}{\kappa^{0.641}+0.875}.
\end{equation}

To conclude this case, we point out that the results reported in Figs. \ref{Fig:LN3} and \ref{Fig:LN4} are exactly the same as those given by the logarithmic negativity. Thus, our results in the three considered cases corroborate that, in fact, $E^{\textrm{cl}}_{\mathcal{N}}$ can be regarded as a classical analog of the logarithmic negativity for a Gaussian state in a system of $N$  coupled harmonic oscillators.

\section{Conclusions}\label{conclusions}
In this article, we consider classical integrable systems and introduce four new classical analogs of quantum quantities that measure entanglement: generalized purities,  R\'enyi entropies,  Bastiaans-Tsallis entropies, and logarithmic negativity. These classical analogous involve an identification of all the action variables of the classical system with a real positive constant, which disappears at the end of the computation. We show through examples that all these classical analogous exactly reproduce the quantum results in the case of Gaussian states. For non-Gaussian states, the results are not reproduced essentially because the covariance matrix does not entirely determine these states.  

In the case of linearly coupled harmonic oscillators, from our results in Sec. \ref{examples}, we observe that the generalized purities are more sensitive than the generalized entropies to the change in the coupling constant of the oscillators, which could be of interest from an experimental point of view. On the other hand, for the Bastiaans-Tsallis and R\'enyi entropies we find that their growth is more negligible for $\alpha >1$ with respect to the von Neumann entropy. But in the case of $\alpha<1$, these entropies are much more sensitive to parameter growth than von Neumann entropy. 

Regarding the chain of generalized oscillators, the results for the generalized entropies and purities are similar to the case of linearly coupled oscillators in the sense that we can reproduce these results only by taking  the corresponding frequencies \eqref{eq:frecuenciasgen}. Furthermore, we have two possibilities for the system to become pure: one implying that the system becomes decoupled, and the other that some self-coupling parameters are imaginary. This last possibility would lead to considering non-Hermitian systems \cite{Bender, Moiseyev}, showing that quantum entanglement has different characteristics for non-Hermitian systems\cite{PTsymmetry, Non-Hermitian}, and this situation already appears in our classical context.

In the case of logarithmic negativity, we observe from the worked-out examples that the classical analog \eqref{LogNegCl2} gives the same results as the quantum case with the advantage that it can be computed rather easily using only classical information. In this sense, our classical analog can reproduce all the quantum information of Gaussian states. We also corroborate this by comparing the predictions of our approach with the results of \cite{Calabrese2012,Calabrese2013} for a finite system, finding an excellent agreement between them. A remarkable result is that, in the case of Gaussian states, the value of the central charge can be obtained from the classical setting. This last point constitutes another advantage of our approach.

In view of these results, our classical analogs could be helpful as a first estimation of quantum effects, and it would be worth extending our results  to non-integrable systems, for instance, to compute generalized fractal dimensions \cite{Zmeskal}.  Furthermore, it will be interesting to generalize these classical analogs to other kinds of states, like non-Gaussian states, perhaps along the lines of \cite{Mattia}, where quantities like purity are expressed in terms of the Wigner function and of which we have an excellent classical analog.

%
%
\section*{Acknowledgments}

This work was partially supported by DGAPA-PAPIIT Grant No. IN105422 and by the grants PID2020-116567GB-C22, CEX2019-000904-S funded by
MCIN/AEI/10.13039/501100011033. Bogar D\'iaz acknowledges support from the CONEX-Plus programme funded by Universidad Carlos III de Madrid and the European Union's Horizon 2020 research and innovation programme under the Marie Sklodowska-Curie Grant Agreement No. 801538.

%
%

%
%

\bibliography{praDGHV}

\end{document}